\documentclass[letterpaper, 9pt, conference]{ieeeconf}  

\IEEEoverridecommandlockouts                              

\usepackage{cite}
\usepackage{ntheorem}
\usepackage{algorithm}
\usepackage{algpseudocode}
\usepackage{subfigure, amsmath, rotating}
\usepackage{amsbsy, amsfonts, graphics, graphicx}
\usepackage{epstopdf}
\usepackage{mathrsfs}
\usepackage{caption}
\usepackage{amssymb}
\usepackage{amsmath}

\usepackage[multiple]{footmisc}
\usepackage{mathtools}

\usepackage{tikz}
\usepackage{float}
\usetikzlibrary{calc,patterns,decorations.pathmorphing,decorations.markings}

\usepackage{textcomp}

\newcommand*{\Scale}[2][4]{\scalebox{#1}{$#2$}}%
\newtheorem{assumption}{Assumption}
\newtheorem{theorem}{Theorem}

\newtheorem{remark}{Remark}
\newtheorem{lemma}{Lemma}

\newtheorem{proposition}{Proposition}

\begin{document}
\title{Real-Time Distributed Model Predictive Control with Limited Communication Data Rates}
\author{Yujia Yang, Ye Wang, Chris Manzie, and Ye Pu
\thanks{Y. Yang, Y. Wang, C. Manzie, and Y. Pu are with the Department of Electrical and Electronic Engineering, University of Melbourne, Parkville VIC 3010, Australia {\tt\small {yujyang1}@student.unimelb.edu.au, \{ye.wang1,manziec,ye.pu\}@unimelb.edu.au}}
\thanks{Y. Yang is supported by the Melbourne Research Scholarship provided by the University of Melbourne.}
\thanks{Y. Wang and Y. Pu acknowledge support from the Australian Research Council via the Discovery Early Career Researcher Awards (DE220100609 and DE220101527), respectively.}
}

\maketitle

\begin{abstract}
The application of distributed model predictive controllers (DMPC) for multi-agent systems (MASs) necessitates communication between agents, yet the consequence of communication data rates is typically overlooked. This work focuses on developing stability-guaranteed control methods for MASs with limited data rates.
Initially, a distributed optimization algorithm with dynamic quantization is considered for solving the DMPC problem.
Due to the limited data rate, the optimization process suffers from inexact iterations caused by quantization noise and premature termination, leading to sub-optimal solutions.
In response, we propose a novel real-time DMPC framework with a quantization refinement scheme that updates the quantization parameters on-line 
so that both the quantization noise and the optimization sub-optimality decrease asymptotically.
To facilitate the stability analysis, we treat the sub-optimally controlled MAS, the quantization refinement scheme, and the optimization process as three interconnected subsystems.
{The cyclic-small-gain theorem is used to derive sufficient conditions on the quantization parameters for guaranteeing the stability of the system under a limited data rate.}
Finally, the proposed algorithm and theoretical findings are demonstrated in a multi-AUV formation control example.
\end{abstract}

\section{Introduction} \label{sec intro}
Model predictive control can systematically handle constraints in determining the control action. In multi-agent systems (MASs) without central decision-makers, distributed model predictive control (DMPC) offers a viable solution.
To apply DMPC, the agents rely on communication networks for solving the underlying distributed optimization problems at every time step.
Thus, it is necessary to consider the communication challenges ubiquitous in communication networks \cite{comcha}, e.g., communication delay, package loss, and limited communication data rate.
In this paper, we focus on implementing DMPC for MASs with a limited data rate and aim to provide stability guarantees for the closed-loop system.

In \cite{Puy}, \cite{magnusson2020maintaining}, distributed optimization algorithms that consider quantization noise arising from limited communication data rates are proposed and provide convergence guarantees when suitable quantization parameters are adopted. 
However, the fixed data rate not only causes inexact iterations but also limits the number of optimization iterations allowed, thus leading to sub-optimal solutions. In turn, the stability of the sub-optimally controlled system may not be guaranteed.

Sub-optimal MPC has been studied extensively, starting with \cite{sub_old}.
In \cite{sub_old1}, \cite{subopt2}, and \cite{sub_old2}, different MPC schemes are combined with real-time stability-guaranteeing mechanisms that are verified on-line. 
In \cite{subopt3} and \cite{dominic1}, the small-gain theorem is used to derive sufficient conditions for the stability of dynamics-optimizer systems, where the former assumes Lipschitz continuity of the MPC problem and the latter considers MPC with only input constraints.
In \cite{9335016}, the small-gain theorem is applied for stability analysis of a MAS interconnected with an output coordinator and nonlinear controllers.
The works \cite{dsubopt1} and \cite{dsubopt2} involve on-line stopping criteria that guarantee the stability of sub-optimal DMPC. 
Despite their efficacy in dealing with sub-optimality, these methods are not readily applicable in DMPC with limited communication data rates.
%
%
There are DMPC schemes designed considering communication data rate.
The work \cite{DMPCwithDATA1} compresses shared data with neural networks to save communication data rate and proves input-to-state-practical stability.
In \cite{DMPCwithDATA2}, a single-chain communication topology is considered and stability is guaranteed with a sufficiently long prediction horizon.
{While the above methods aim to reduce the communication data rate required for solving the DMPC problem, they do not explicitly consider the effect of the limited data rate on the closed-loop stability.}

\begin{figure}[t!]
 \centering
\includegraphics[width=1\hsize]{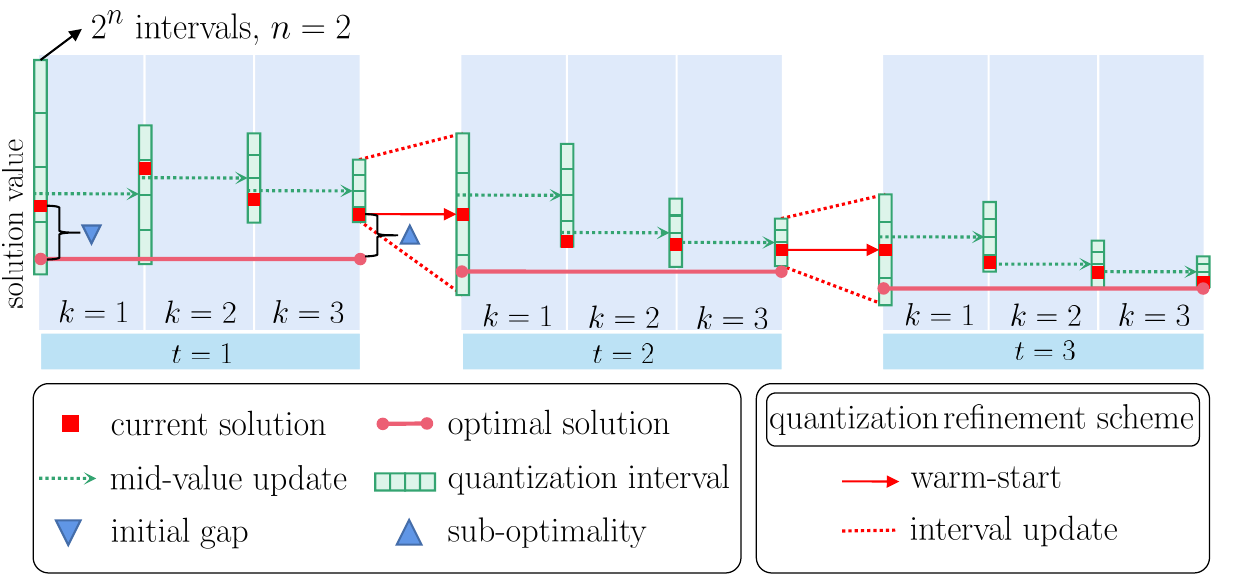}
\caption{The on-line quantization refinement scheme.}
\label{intersection}
\end{figure}

{In this work, we propose a DMPC framework for controlling a MAS under a limited data rate, with explicitly guaranteed closed-loop stability.
Our previous work \cite{ACC2021} is an initial step in this direction, which solves DMPC problems using the distributed optimization algorithm in \cite{Puy} with quantization parameters determined off-line considering a limited data rate.}
{Compared to \cite{ACC2021}, we refine the quantization parameters on-line to adapt to the change in sub-optimality and use a different stability analysis method, which allows for stronger stability guarantees.
Specifically, we make the following main contributions:}
\begin{itemize}
    \item {We propose a novel real-time DMPC framework with a quantization refinement scheme.
    The scheme includes
an off-line stage that designs initial quantization parameters given a limited communication data rate,
and an on-line stage (illustrated in Fig. 1) that implements the quantization refinement scheme, the optimization process, and the DMPC controller.} 
\item {Given a limited data rate, we derive sufficient, off-line-determined conditions on the quantization parameters such that the system controlled by the proposed framework is recursively feasible and stable.}
{Specifically, we consider the sub-optimally controlled MAS, the quantization refinement scheme, and the optimization process as three interconnected subsystems.}
{This formulation facilitates stability analysis of the closed-loop system using the cyclic-small-gain theorem, leading to input-to-state stability (ISS) guarantee w.r.t a quantization refinement parameter} {that associates with state estimation error.}
\end{itemize}

\section{Preliminaries}\label{sect prem}
For an optimization problem and an optimization algorithm, let $z^*$ and $z^k \in \mathbb{R}^{m_z} $ denote the optimal and $k_\text{th}$-iteration solutions, respectively. 
Let the subscript $t$ denote the discrete time step of a controlled system.
Let $I$ denote the identity matrix and Id denote the identity function. 
Let $\mathbb{S}^n_{++}$ denote the set of all positive definite matrices with size $n \times n$.
{For a vector $x \in \mathbb{R}^{m_x}$ and a matrix $H \in \mathbb{S}^{m_x}_{++}$,
let $\|x\|_{\infty}$, $\|x\|$, and $\|x\|_H$ denote the $\infty$-norm, $2$-norm, and the weighted 2-norm of $x$, respectively.
Note that $\|x\|_{\infty} \leq \|x\| \leq \sqrt{m_x} \|x\|_{\infty}$.}
Let $\| H \|$ denote the induced norm of $H$.
Let $\underline{\lambda} (H )$ and $\overline{\lambda} (H) $ denote the smallest and largest eigenvalues of $H$, respectively.
For a continuous function $\alpha(s)$, $\alpha(s) \in \mathcal{K}$ if $ \alpha(0) = 0 $ and $ \dot{\alpha}(s) > 0 $.
Furthermore, $\alpha(s) \in \mathcal{K}_{\infty}$ if $\alpha(s)$ is unbounded.
{Given functions $\alpha_1(w): \mathcal{W} \rightarrow \mathcal{Y}$ and $\alpha_2(y): \mathcal{Y} \rightarrow \mathcal{Z}$, we define $\alpha_1 \circ \alpha_2 (w) = \alpha_2(\alpha_1 (w) ) : \mathcal{W} \rightarrow \mathcal{Z}$.}
The operators $ \left \lceil  \cdot \right \rceil$ and $ \left \lfloor  \cdot \right \rfloor$ denote the ceil and floor operations, respectively.
The operator $\operatorname{blkdiag}(H_1,\cdots,H_m)$ yields a matrix $H$ containing the matrices $H_1,\cdots,H_m$ in its diagonal.
We denote the projection of a vector $z \in \mathbb{R}^{m_{z}}$ onto the set $\mathcal{C} \subseteq \mathbb{R}^{m_{z}} $ as $\operatorname{Proj}_{\mathcal{C}}(z):=\operatorname{argmin}_{\bar{z} \in \mathcal{C}}\|\bar{z}-z\|$.
Let the uniform quantizer be defined as $Q(x):=\bar{x}+\operatorname{sgn}(x-\bar{x}) \cdot \frac{l}{2^n}\cdot\left\lfloor\frac{\|x-\bar{x}\|}{{l}/{2^n}}+\frac{1}{2}\right\rfloor$, with mid-value $\bar{x}$, quantization interval $l$, and bit number $n$.
The cyclic-small-gain theorem presented below is relevant to the stability analysis:

\begin{lemma}[Lemma 1, \cite{smallgain}]
Consider an interconnected system composed of $M$ subsystems with discrete dynamics
\begin{equation}
    x_{i,t+1} = f_i(x_{i,t},x_{j,t},u_{i,t}), \; \forall i,j \in \{1,\cdots,M\}, i \neq j,
\end{equation}
where $x_{i,t}$, $x_{j,t}$, and $u_{i,t}$ denote the states, interconnected states, and external inputs, respectively.
Suppose each subsystem admits a continuous ISS-Lyapunov function satisfying
\begin{align}
      \tilde{\alpha}_{i,l} \left(\|x_{i,t}\|\right) & \leq  V_{i}\left(x_{i,t}\right) \leq \tilde{\alpha}_{i,u} \left( \|x_{i,t}\| \right), \\
  V_i(x_{i,t+1}) & -  V_i(x_{i,t}) \leq - \tilde{\alpha}_i (V_i(x_{i,t}))  \nonumber  \\ 
    &  + \max_{j \in \{1,\cdots,M \},j \neq i}  \left( \tilde{\gamma}_{i,j} ( V_j(  x_{j,t} )) ,\tilde{\gamma}_i^e( \|  x^e_{i,t} \|) \right), \label{lyp decreasing condition}
\end{align}
where $\tilde{\alpha}_{i,l}, \tilde{\alpha}_{i,u}, \tilde{\alpha}_i \in \mathcal{K}_{\infty}$, $\left(\mathrm{Id}- \tilde{\alpha}_{i}\right) \in \mathcal{K}$ and $\tilde{\gamma}_{i,j}, \tilde{\gamma}_i^e \in \mathcal{K} \cup\{0\}$.
Then, the interconnected system is ISS with $u_{t} = [u_{1,t}^\top, \allowbreak \cdots, \allowbreak u_{M,t}^\top]^\top$ as the input,
if there exist positive definite functions $\tilde{\mu}_{i}$ satisfying $ \mathrm{Id}-\tilde{\mu}_{i} \in \mathcal{K}_{\infty}$ such that the ISS gain functions $\mathcal{X}_{i,j} :=\tilde{\alpha}_{i}^{-1} \circ\left(\mathrm{Id}-\tilde{\mu}_{i}\right)^{-1} \circ \tilde{\gamma}_{i,j}$ satisfy 
\begin{align} \label{small gain conditions original}
    \mathcal{X}_{j,j'} \circ \mathcal{X}_{j',j''} \circ  \cdots \circ \mathcal{X}_{r,j}< \textup{ Id }
\end{align}
for each $r = \{2,\cdots,M\}$ and all $ 1 \leq j < j' < j'' < r$, i.e., for all the interconnected cycles in the system.
\end{lemma}

\section{Distributed Optimization with Quantization}

{We solve a class of DMPC problems over a system of $M$ agents that communicate synchronously under limited communication data rates via an undirected graph $\mathcal{G} = (\mathcal{M},\mathcal{E})$, with vertex set $\mathcal{M} = \{1,\cdots,M\}$ and edge set $\mathcal{E} \subseteq \mathcal{M} \times \mathcal{M}$.
If $(i,j) \in \mathcal{E}$, then agent $i$ and $j$ are neighbours and communicate with each other.
Let $\mathcal{N}_{i} = \{j \mid (i,j) \in \mathcal{E}\}$ denote the set of neighbors of agent $i$, including agent $i$ itself.
Let $d$ denote the degree of $\mathcal{G}$.
The DMPC problems can be formulated as the following distributed optimization problem:
    \begin{align}  \label{problem:PDproblem}
         \min _{z}  \; & f\left(z \right)  =\sum_{i \in \mathcal{M }}  f_{i}\left(z_{\mathcal{N}_{i}} \right), \\
       \text{s.t. }   &  z_{i} \in \mathcal{C}_{i},  z_{\mathcal{N}_{i}}  = \;  E_{i}  z,   z_{j} = F_{ij} z_{\mathcal{N}_{i}},   \nonumber
    \end{align}    
where $z=[ z_{1}^\top, \cdots, z_{M}^\top ]^\top \in \mathbb{R}^{m_z}$ and $z_i \in \mathbb{R}^{m_{z_i}}$ denote the global and local decision variables, respectively.
Let $E_i \in \mathbb{R}^{m_{z_i} \times m_{z}} $, 
$F_{ji} \in \mathbb{R}^{{m_{z_{\mathcal{N}_i}}} \times m_{z}} $ be the selection matrices defined w.r.t. $\mathcal{N}_i$.
Let $z_{\mathcal{N}_{i}}$ be the stacked decision variables of the agents in ${\mathcal{N}_{i}}$.
Let $\bar{m} := \max_{i \in \mathcal{M}} m_{z_i}$ denote the largest size of local decision variables.
Let $\mathcal{C}_i$ be the local constraint set.}
\begin{assumption}
    All the local constraint sets {$\mathcal{C}_i$} are convex.
\end{assumption}
\begin{assumption} \label{ap1}
All local cost functions $f_i(\cdot)$ are strongly convex with Lipschitz continuous gradients $\nabla f_i(\cdot)$ satisfying $\| \nabla  f_i (z_{i,t_1}) - \nabla  f_i (z_{i,{t_2}})\| \leq L_i \|z_{i,t_1} - z_{i,t_2} \| $ for any $z_{i,t_1}$ and $z_{i,t_2}$.
\end{assumption}
Assumption 2 implies the global cost function $f(\cdot)$ is strongly convex with a convexity modulus $\alpha_{f}$ and a Lipschitz continuous gradient $\nabla f(\cdot)$ satisfying $\| \nabla  f (z_{t_1}) - \nabla  f (z_{t_2})\| \leq L_f \|z_{t_1} - z_{t_2} \|$.
%

\begin{algorithm}[t]
\caption{Distributed Optimization with Dynamic Quantization}
\begin{algorithmic}[0]
    \State \textbf{\textup{Require:}} Initialize $ \check{z}^{-1}_i =  z^{0}_i $ $\check{\nabla}f^{-1}_{i}=\nabla f_{i}(\operatorname{Proj}_{\mathcal{C}_{\mathcal{N}_{i}}}(z_{\mathcal{N}_{i}}^{0}))$, {$C_i$}, $K$, $(1-\gamma)<\rho<1$, $\gamma = \frac{\alpha_f}{L_f}$, and $ \eta<\frac{1}{L_f}$;
    \For{$k = 0,1,\cdots,K$}
    \For{$\text{agent } i \in \mathcal{M}$} \Comment in parallel
        \State 1. Update {$l_{i}^{k}= \rho^{k} C_{i}$ and $\bar{z}_{ i}^{k}=\check{z}_{i}^{k-1}$ for $Q_{\alpha, i}^{k}$;}
\State 2. Quantize local variable: $\check{z}_{i}^{k}=Q_{\alpha,i}^{k}\left(z_{i}^{k}\right)$;
\State 3. Send $\check{z}^{k}_{i}$ to agent $j$ for all $j\in\mathcal{N}_i$;
\State 4. Compute the projection: $ \tilde{z}_{\mathcal{N}_{i}}^{ k}=\operatorname{Proj}_{\mathcal{C}_{\mathcal{N}_{i}}}\left(\check{z}_{\mathcal{N}_{i}}^{k}\right)$;
\State 5. Compute $\nabla f^{k}_{i} = \nabla f_i \left(\tilde{z}_{\mathcal{N}_{i}}^{k}
\right)$;
\State 6. Update {$ l_{i}^{k}= \rho^{k} C_{i} $ and $\bar{\nabla} f_{i}^{k}=\check{\nabla} f_{i}^{k-1}$ for $Q_{\beta, i}^{k} $;}
\State 7. Quantize local gradient: $\check{\nabla} f_{i}^{k}=Q_{\beta, i}^{k}\left(\nabla f_{i}^{k}\right)$;
\State 8. Send $\check{\nabla} f^{k}_{i}$ to agent $j$ for all $j\in\mathcal{N}_i$;
\State 9. $z^{k+1}_{i} = \mathrm{Proj}_{\mathcal{C}_{i}} \left(z^{k}_{i} - \eta \sum_{j \in \mathcal{N}_{i}} F_{j i} \check{\nabla} f_{j}^{k} \right)$; 
    \EndFor
    \EndFor
\end{algorithmic}
\end{algorithm}

{To solve \eqref{problem:PDproblem}, we consider Alg. 1, modified from the distributed optimization algorithm with dynamic quantization proposed in \cite{Puy}.}
Uniform quantizers $Q_{\alpha, i}^{k}$ and $Q_{\beta, i}^{k}$ are used by agent $i$ for transmitting local decision variables $z^k_i$ and gradients $\nabla f^{k}_{i}$ at iteration $k$, respectively.
They are parametrized by the quantization bit number $n$, {quantization interval $l^{k}_{i}$, and mid-values $\bar{z}_i^K$ and $\bar{\nabla} f_{i}^{k}$, respectively.}
{The quantization interval is progressively updated according to $ l^{k}_i= C_i \rho^{k}$, with base quantization interval $C_i$ and the shrinkage constant $\rho \in ( 1 - \gamma,1 )$, $\gamma = {\alpha_f}/{L_f}  \in (0, 1)$.}
The mid-values are updated according to $\bar{z}_{ i}^{k}=\check{z}_{i}^{k-1}$ and $\bar{\nabla} f_{i}^{k}=\check{\nabla} f_{i}^{k-1}$.
{The projections $\operatorname{Proj}_{\mathcal{C}_{\mathcal{N}_{i}}}$ and $\operatorname{Proj}_{\mathcal{C}_{i}}$ ensure the quantized variable $\check{z}_{\mathcal{N}_{i}}^{k}$ and the local solution computed from quantized gradient $\check{\nabla} f_{j}^{k}$ are feasible despite the quantization noises, for all iteration $k \geq 0$.
The constraint set $\mathcal{C}_{\mathcal{N}_{i}} = \mathcal{C}_{i} \times \cdots  \times  \mathcal{C}_{j}$ represents the constraints of all the agents in $\mathcal{N}_{i}$.}
Let $\eta$ denote the optimization step-size.
Each blue block in Fig. 1 corresponds to the implementation of Alg. 1 for $n = 2$ and $K = 3$.

{The main difference between Alg. 1 and \cite{Puy} is we initialize the local quantizers $Q_{\alpha, i}^{k}$ and $Q_{\beta, i}^{k}$ of agent $i$ with a local base quantization interval $C_i$ instead of a global one.
This change eases the need for the agents to agree on a  base quantization interval through centralized communication.
Correspondingly, we propose a new condition on the base quantization interval $C_{i}$ and the bit number $n$, and a new convergence bound for Alg. 1.}

\begin{lemma} \label{tm1}
{Consider Alg. 1 applied to solve problem \eqref{problem:PDproblem}.
Suppose {Assumptions 1 and 2} hold. 
Let $z^0$ be the initial solution and ${\Delta z^{0}} := z^{0}-z^{*}$ be the initial gap.
If the quantization bit number $n$ satisfies 
\begin{align} \label{n definition}
   2^n  \rho - \sqrt{d \bar{m}} > 0,
\end{align}
and the local base quantization interval $C_i$ satisfies
\begin{equation} \label{C condition}
     a_{1} \| {\Delta z^{0}} \| + a_2 C_{\Sigma} \leq C_i, \;i \in \mathcal{M},
\end{equation}
then the sub-optimality ${\Delta z^{K}} := z^{K}-z^{*}$ generated by {Alg. 1} satisfies
\begin{align} \label{subbound}
    \| {\Delta z^{K}} \|   \leq \rho^{K} ( \| {\Delta z^{0}} \| + a_3 C_{\Sigma} ),
\end{align}
where $C_{\Sigma} := \sum_{i \in \mathcal{M}} C_i$ and
\begin{equation} \label{a_1_3 definitions}
  \begin{aligned}
    a_1  := &  \max (  \frac{ 2^{n+1} \max_{i \in \mathcal{M}} (L_i) (1+\rho)}{ 2^n  \rho -   \sqrt{d \bar{m}} }, {2(1+\rho) }/{\rho}  ),  \allowdisplaybreaks \\
    a_2 := &  \max (  \frac{ \max_{i \in \mathcal{M}}(L_i) (1+\rho) (2^{n+1} a_3 + { \sqrt{\bar{m}}  } ) }{   2^n  \rho -   \sqrt{d \bar{m}} },   \allowdisplaybreaks \\
    & \qquad\;\;\;  ({   2^{n+1} (1+\rho) a_3
     + \sqrt{\bar{m}} })/({  2^{n} \rho})  ),    \allowdisplaybreaks \\
      a_3 := & \frac{  \sqrt{ \bar{m}} ( M \max_{i \in \mathcal{M}} (L_i)  +\sqrt{L_f d}  + \sqrt{d}   )    \rho}{L_f(\rho+\gamma-1)(1-\gamma) 2^{n+1} }.
\end{aligned}  
\end{equation}
\begin{proof} 
The proof contains two parts:
\textbf{(i)} inspired by Lemma 3.10 \cite{Puy}, we develop a one-step bound on $\|{\Delta z^{p+1}}\|$ in the form of \eqref{subbound} which holds if the quantization errors are bounded at iteration $p$.
\textbf{(ii)} inspired by Lemma 3.17 \cite{Puy}, we use induction to show the quantization errors are bounded at every iteration $k$ and the one-step convergence bound holds at every iteration $k$.

\textbf{(i)} We first construct the one-step bound.
For a uniform quanitzer $Q(z)$ with quantization interval $l$ and bit number $n$, if the input $z$ falls inside the quantization interval of $Q(z)$, then it holds that
\begin{align}
 \| z - Q(z)\| \leq \frac{l}{2^{n+1}}. \label{quant error bound}
\end{align}
When $k = p$, if $z_i^p$ and $\nabla f_i^p$ fall inside the quantization intervals of $Q_{\alpha, i}^{p}$ and $Q_{\beta, i}^{p}$, respectively, then the quantization errors $\alpha_i^p=\check{z}_i^p-z_i^p$ and $\beta_i^p=\check{\nabla} f_i^p-\nabla f_i^p$ satisfy 
\begin{align}
\left\|\alpha_i^p\right\| & \leq \sqrt{m_i}  \left\|\alpha_i^p\right\|_{\infty} \overset{\eqref{quant error bound}}{\leq} \sqrt{\bar{m}} \frac{l_{i}^p}{2^{n+1}} = \sqrt{\bar{m}} \frac{\rho^p C_i}{2^{n+1}},  \label{error bound 1 v0} \\
\left\|\beta_i^p\right\| & \leq \sqrt{\sum_{j \in \mathcal{N}_i} m_i}  \left\|\beta_i^p\right\|_{\infty}  \overset{\eqref{quant error bound}}{\leq} \sqrt{d \bar{m}}   \frac{l_{i}^p}{2^{n+1}} = \sqrt{d \bar{m}} \frac{\rho^p C_i}{2^{n+1}}, \label{error bound 2 v0}
\end{align}
respectively.
From Lemma 3.8 \cite{Puy}, the error sequences from Alg. 1,
\begin{align*}
    \mathrm{e}^p=\sum_{i \in \mathcal{M}} E_i^T \nabla f_i\left(\tilde{z}_{\mathcal{N}_i}^p\right)+\sum_{i \in \mathcal{M}} E_i^T \beta_i^p-\sum_{i \in \mathcal{M}} E_i^T \nabla f_i\left({\mathcal{N}_i}^p\right)
\end{align*}
and $\epsilon^p=\frac{1}{2}\|z^p-\tilde{z}^p\|^2$ satisfy 
\begin{align}
    \|\mathrm{e}^p\| \leq & \sum_{i \in \mathcal{M}} L_i \sum_{j \in \mathcal{N}_i}\|\alpha_j^p\|+\sum_{i \in \mathcal{M}}\|\beta_i^p\|   \label{error bound 1 v1}\\
    \sqrt{\epsilon^p} \leq &\frac{\sqrt{2}}{2} \sum_{i \in \mathcal{M}}\|\alpha_i^p\|, \label{error bound 2 v1}
\end{align}
respectively.
Then, applying the bounds in \eqref{error bound 1 v0} and \eqref{error bound 2 v0} to \eqref{error bound 1 v1} and \eqref{error bound 2 v1}, we obtain
\begin{align}
     \left\|\mathrm{e}^p\right\| & \leq \frac{\sqrt{\bar{m}}  }{2^{n+1}}  ( \max_{i \in \mathcal{M}} (L_i) \sum_{i \in \mathcal{M}}\sum_{j \in \mathcal{N}_i} C_{j} + \sqrt{d} C_{\Sigma} ) \rho^p \leq C_a \rho^p, \label{error bound 1 v2} \\
     \sqrt{\epsilon^p} &  \leq  \frac{\rho^p \sqrt{2d \bar{m}} }{2^{n+2}}  C_{\Sigma} = C_b \rho^p, \label{error bound 2 v2}
\end{align}
where $C_a = \frac{\sqrt{\bar{m}}  }{2^{n+1}}  ( \max_{i \in \mathcal{M}} (L_i) M  C_{\Sigma} + \sqrt{d} C_{\Sigma} ) $
and 
$C_b = \frac{\sqrt{2d \bar{m}} }{2^{n+2}}  C_{\Sigma}$.
In Proposition 2.5 of \cite{Puy}, it is proven that 
\begin{align} \label{hid inter 0}
    \| z^{p+1}-z^* \| \leq(1-\gamma)^{p+1}   ( 
 \|z^0-z^* \|+\Gamma^p ),
\end{align}
where 
\begin{align} \label{hid inter 1}
    \Gamma^p=\sum_{p=0}^p(1-\gamma)^{-q-1}  \left(\frac{1}{L_f} \|\mathrm{e}^q \|+\sqrt{\frac{2}{L_f}} \sqrt{\epsilon^q} \right)
\end{align}
accounts for the inexactness in an iteration of the inexact proximal-gradient method.
Plugging \eqref{hid inter 1} into the r.h.s of \eqref{hid inter 0}, using \eqref{error bound 1 v2} and \eqref{error bound 2 v2} to bound $\left\|e^p\right\|$ and $\sqrt{\epsilon^p}$, respectively, and using the fact that $0 < (1-\gamma)<\rho<1$, we obtain
\begin{align}
& \| \Delta z^{p+1} \| = \| z^{p+1}-z^* \| \nonumber \\
\leq & (1-\gamma)^{p+1} \| \Delta z^{0} \| \\
& +\frac{\left(C_a+\sqrt{2 L_f} C_b\right)}{L_f} \sum_{q=0}^p \rho^q(1-\gamma)^{p+1-q-1} \\
\leq & \rho^{p+1}\left( \| \Delta z^{0} \| +\frac{(C_a+\sqrt{2 L_f} C_b)}{L_f(1-\gamma)} \sum_{q=0}^p\left((1-\gamma) / \rho\right)^{p+1-q}\right) \label{sum term 1} \\
 \leq &\rho^{p+1}\left( \| \Delta z^{0} \| +\frac{(C_a+\sqrt{2 L_f} C_b)}{L_f(1-\gamma)}  \frac{1-\left( (1-\gamma) / \rho \right)^{p+1}}{1- (1-\gamma) / \rho}\right) \label{sum term 2} \\
 \leq & \rho^{p+1}\left( \| \Delta z^{0} \| +\frac{(C_a+\sqrt{2 L_f} C_b) \rho}{L_f(1-\gamma)(\rho+\gamma-1)}\right)  \label{sum term 3}\\
 = & \rho^{p+1}( \| \Delta z^{0} \| + a_3 C_{\Sigma}), \label{one step bound}
\end{align}
where the property of geometric series is used to obtain \eqref{sum term 2} by bounding $\sum_{q=0}^p\left((1-\gamma) / \rho\right)^{p+1-q}$ in \eqref{sum term 1}.
The term $a_3$ in \eqref{one step bound}, which is defined in \eqref{a_1_3 definitions}, is obtained by plugging $C_a$ and $C_b$ into \eqref{sum term 3} and carrying out simplification.

\textbf{(ii)} We prove by induction that $z_i^k$ and $\nabla f_i^k$ fall inside the quantization intervals of $Q_{\alpha, i}^{k}$ and $Q_{\beta, i}^{k}$, respectively, for all $k \geq 0$, so that the one-step bound in (23) holds at every step.

Base case: When $k = 0$, we initialize $C_i > 0$ and $\check{z}_i^{-1} = z_i^0 = 0$ $\forall i \in \mathcal{M}$, which satisfies 
\begin{align}
   & \|z_i^0-\bar{z}_{ i}^0\|_{\infty}=\|z_i^0-\check{z}_i^{-1}\|_{\infty}=0 \leq \frac{l_{i}^0}{2}=\frac{C_i}{2}, \label{base case a} \\
  &  \|\nabla f_i^0-\bar{\nabla} f_{ i}^0\|_{\infty}=\|\nabla f_i^0-\check{\nabla} f_i^{-1}\|_{\infty} \nonumber \\ 
  & = \|\nabla f_i(\tilde{z}_{\mathcal{N}_i}^0)-\nabla f_i(\operatorname{Proj}_{\mathcal{C}_{\mathcal{N}_{i}}}(0))\|=0 \leq \frac{l_{i}^0}{2}=\frac{C_i}{2}. \label{base case b}
\end{align}
Thus, bounded quantization errors imply the input variable are bounded by the quantization intervals of the quantizers.

Induction case: When $k = p \geq 0$, suppose $\|z_i^k-\bar{z}_{ i}^k\|_{\infty} \leq {l_{ i}^k}/{2} $ and $\|\nabla f_g^k-\bar{\nabla} f_{i}^k\|_{\infty} \leq {l_{i}^k}/{2}$ for $ 0 \leq k \leq p$.
We need to prove
\begin{align}
  & \|z_i^{p+1}-\bar{z}_{ i}^{p+1}\|_{\infty} \leq {l_{ i}^{p+1}}/{2}, \label{induction a v0} \\
 &  \|\nabla f_g^{p+1}-\bar{\nabla} f_{i}^{p+1}\|_{\infty} \leq {l_{i}^{p+1}}/{2}. \label{induction b v0}
\end{align}

We first prove \eqref{induction a v0}. The quantization error from Alg. 1 satisfies
\begin{align}
    & \|z_i^{p+1}-\bar{z}_{i}^{p+1}\|_{\infty}  = \|z_i^{p+1}-\check{z}_{i}^p\|_{\infty} \leq  \|z^{p+1}-\check{z}^p\|_{\infty} \allowdisplaybreaks  \\
    = & \|z^{p+1}-z^p-\sum_{i \in \mathcal{M}} E_i^T F_{i i}^T \alpha_i^p\|_{\infty} \allowdisplaybreaks  \\
  \leq  &  \|z^{p+1}-z^p\|_{\infty}+\|\sum_{i \in \mathcal{M}} E_i^T F_{i i}^T \alpha_i^p\|_{\infty} \allowdisplaybreaks  \\
    \leq  & \|z^{p+1}-z^{\star}\|_{\infty}+\|z^p-z^{\star}\|_{\infty}+\|\sum_{i \in \mathcal{M}} E_i^T F_{i i}^T \alpha_i^p\|_{\infty} \allowdisplaybreaks  \\
     \leq   & \| \Delta z^{p+1}\|+\| \Delta z^{p} \|+\sum_{i \in \mathcal{M}}\|\alpha_i^p\|, \label{induction mid v0}
\end{align}
where we used $\| E_i\| = \| F_{ij} \| = 1$ to obtain \eqref{induction mid v0}.
Since $\|z_i^k-\bar{z}_{ i}^k\|_{\infty} \leq {l_{ i}^k}/{2} $ and $\|\nabla f_g^k-\bar{\nabla} f_{i}^kp\|_{\infty} \leq {l_{i}^k}/{2}$ for $ k = p$ and $k = p-1$, we can bound $\| \Delta z^{p+1}\|$ and $\| \Delta z^{p}\|$ with \eqref{one step bound} and use \eqref{error bound 1 v0} to obtain 
\begin{align}
   & \|z_i^{p+1}-\bar{z}_{i}^{p+1}\|_{\infty} \nonumber \\
    \leq  & \rho^{p+1}(\| \Delta z^{0}\| + a_3 C_{\Sigma})   +\rho^{p}(\| \Delta z^{0}\| + a_3 C_{\Sigma}) \nonumber
    \\ & + \rho^p \frac{ \sqrt{\bar{m}}  }{2^{n+1}}  C_{\Sigma}, \\
  \leq &  \frac{\rho^{p+1}}{2} ( a_1 \| \Delta z^{0}\| + a_2  C_{\Sigma} )\label{induction mid v1}
\end{align}
where \eqref{induction mid v1} holds from the definitions of $a_1$, $a_2$ in \eqref{a_1_3 definitions}, and the condition \eqref{n definition}.
Then, condition \eqref{C condition} implies \eqref{induction a v0} holds for all $k = p \geq 0$.

We then prove \eqref{induction b v0}. From Alg. 1, the gradient satisfies
    \begin{align}
&\|\nabla f_i^{p+1}-\bar{\nabla} f_{ i}^{p+1}\|_{\infty}  =\|\nabla f_i^{p+1}-\check{\nabla} f_i^p\|_{\infty} \\
 = &\|\nabla f_i\left(\tilde{z}_{\mathcal{N}_i}^{p+1}\right)-\nabla f_i\left(\tilde{z}_{\mathcal{N}_i}^p\right)+\beta_i^p\|_{\infty} \\
\leq &  \|\nabla f_i\left(\tilde{z}_{\mathcal{N}_i}^{p+1}\right)-\nabla f_i\left(\tilde{z}_{\mathcal{N}_i}^p\right)\|+\|\beta_i^p\| \\
 \leq & L_i\|\tilde{z}_{\mathcal{N}_i}^{p+1}-\tilde{z}_{\mathcal{N}_i}^p\|+\|\beta_i^p\| \\
 \leq & L_i ( \|z_{\mathcal{N}_i}^{p+1}-z_{\mathcal{N}_i}^p\|+ \|\tilde{z}_{\mathcal{N}_i}^{p+1}-z_{\mathcal{N}_i}^{p+1}\|+ \|\tilde{z}_{\mathcal{N}_i}^p-z_{\mathcal{N}_i}^p\| )+\|\beta_i^p\|. \label{induction mid v2}
\end{align}
Since $\check{z}_{\mathcal{N}_i}^p, \check{z}_{\mathcal{N}_i}^{p+1} \in \mathcal{C}_{\mathcal{N}_{i}}$, we have from Lemma 3.7 \cite{Puy} that 
\begin{align}
  &  \|\tilde{z}_{\mathcal{N}_i}^p-z_{\mathcal{N}_i}^p\| = \| \operatorname{Proj}_{\mathcal{C}_{\mathcal{N}_{i}}}(\check{z}_{\mathcal{N}_i}^p )-z_{\mathcal{N}_i}^p\| \leq \|\check{z}_{\mathcal{N}_i}^p-z_{\mathcal{N}_i}^p\|, \label{proj 1} \\
  &  \|\tilde{z}_{\mathcal{N}_i}^{p+1}-z_{\mathcal{N}_i}^{p+1}\| = \|\operatorname{Proj}_{\mathcal{C}_{\mathcal{N}_{i}}}(\check{z}_{\mathcal{N}_i}^{p+1})-z_{\mathcal{N}_i}^{p+1}\| \leq\|\check{z}_{\mathcal{N}_i}^{p+1}-z_{\mathcal{N}_i}^{p+1}\|. \label{proj 2}
\end{align}
Applying the above bounds to \eqref{induction mid v2}, we obtain
\begin{align}
   & \|\nabla f_i^{p+1}-\bar{\nabla} f_{ i}^{p+1}\|_{\infty} \nonumber \\
& \leq L_i(\|z_{\mathcal{N}_i}^{p+1}-z_{\mathcal{N}_i}^p\|+ \|\check{z}_{\mathcal{N}_i}^{p+1}-z_{\mathcal{N}_i}^{p+1}\|+\|\check{z}_{\mathcal{N}_i}^p-z_{\mathcal{N}_i}^p\|)+\|\beta_i^p\| \\
& \leq L_i\|z_{\mathcal{N}_i}^{p+1}-z_{\mathcal{N}_i}^p\|+L_i \sum_{j \in \mathcal{N}_i}(\|\alpha_j^{p+1}\|+\|\alpha_j^p\|)+\|\beta_i^p\| \\
& \leq L_i\|z^{p+1}-z^p\|+L_i \sum_{j \in \mathcal{N}_i}(\|\alpha_j^{p+1}\|+\|\alpha_j^p\|)+\|\beta_i^p\| \\
& \leq \max_{i \in \mathcal{M}} (L_i) ( \| \Delta z^{p+1} \|+\| \Delta z^{p} \| \nonumber \\
& \quad + \sum_{j \in \mathcal{N}_i}(\|\alpha_j^{p+1}\|+\|\alpha_j^p\|))+\|\beta_i^p\|. \label{induction mid v3}
\end{align}
Due to the induction assumption, we can bound $\| \Delta z^{p+1} \|$ and $\| \Delta z^{p} \|$ in \eqref{induction mid v3} with \eqref{one step bound}, and then use \eqref{error bound 1 v0} and \eqref{error bound 2 v0}, to obtain 
\begin{align}
   & \|\nabla f_i^{p+1}-\bar{\nabla} f_{ i}^{p+1}\|_{\infty} \nonumber \\
 \leq & \max_{i \in \mathcal{M}} (L_i)  \rho^{p+1}(\| \Delta z^{0} \| + a_2 C_{\Sigma}) + \rho^{p}(\| \Delta z^{0} \| + a_2 C_{\Sigma}) \nonumber \\
 & + \max_{i \in \mathcal{M}} (L_i) \sum^M_{i  = 1}(\|\alpha_i^{p+1}\|+\|\alpha_i^p\|)+\|\beta_i^p\| \\
\leq & \rho^{p} \max_{i \in \mathcal{M}} (L_i) (1+\rho) (\| \Delta z^{0} \| + a_2 C_{\Sigma})  \nonumber \\
 & + \frac{ \rho^{p} \max_{i \in \mathcal{M}} (L_i)  \sqrt{\bar{m}}  }{2^{n+1}}  (1 + \rho) C_{\Sigma}  +   \frac{ \rho^p \sqrt{d \bar{m}}  }{2^{n+1}}  C_{i} \\
 \leq  & \frac{\rho^{p+1}}{2} ( a_1(n)\| \Delta z^{0} \| + a_2(n) C_{\Sigma} )\label{induction mid 2 v1}
\end{align}
where \eqref{induction mid 2 v1} holds from the definitions of $a_1$, $a_2$ in \eqref{a_1_3 definitions}, and the condition \eqref{n definition}.
Then, condition \eqref{C condition} implies \eqref{induction b v0} holds for all $k = p \geq 0$.

Finally, we combine the base case and the induction case to conclude (26) and (27) hold for all $k \geq 0$, i.e., the values $z^k_i$ and $\nabla f^k_i$ generated by Alg. 1 always fall inside the quantization intervals of $Q_{\alpha, i}^{k}$ and $Q_{\beta, i}^{k}$, respectively.
As a result, the solution $z^k$ generated by Alg. 1 satisfies \eqref{one step bound} for all $k \geq 0$, i.e., \eqref{subbound} holds.
\end{proof}}
\end{lemma}


\section{Real-Time DMPC Framework with Quantization Refinement} \label{sec framework}

In this section, we propose a novel real-time DMPC framework with a quantization refinement scheme. 
We first introduce the DMPC problem we aim to solve under a limited communication data rate.

\subsection{DMPC with Coupled Cost Functions}
Consider a MAS with $M$ agents whose dynamics are 
\begin{equation}  \label{local dynamics} 
x_{i,t+1} = A_i x_{i,t}+ B_i u_{i,t}, \:\: \forall i  \in \mathcal{M},  
\end{equation}
with states $x_{i,t} \in \mathbb{R}^{m_{x_i}}$, inputs $u_{i,t} \in \mathbb{R}^{m_{u_i}} $, and $(A_i,B_i)$ being controllable.
{Agent $i$ can communicate locally with neighbouring agents in $\mathcal{N}_i$.}
The global dynamics of the MAS is 
\begin{equation}  \label{global dynamics} 
    {{X}_{t+1} = A {X}_{t}+ B {U}_{t},}
\end{equation}
{where $A = \operatorname{blkdiag}(A_1,\cdots,A_M)$, $B = \operatorname{blkdiag}(B_1,\cdots,B_M)$, $X_t = [x_{1,t}^\top,\cdots,x_{M,t}^\top]^\top$, and $U_t = [u_{1,t}^\top,\cdots,x_{M,t}^\top]^\top$.
We stabilize \eqref{global dynamics} to the origin using the following DMPC formulation:}
\begin{subequations} \label{DMPC problem}
\begin{align}
&  (\textbf{X}^*_{t}, \textbf{U}^*_{t}) = \underset{{\mathbf{x}_{i,t},  \mathbf{u}_{i,t}}}{\operatorname{argmin}} \sum_{i=1}^{M} \left(  \sum_{\tau=0}^{N-1} \| [ \mathrm{x}^\top_{\mathcal{N}_i ,t}(\tau), \mathrm{u}^\top_{\mathcal{N}_i,t}(\tau) ]^\top \|_{H_{i}} \right) \nonumber \\
& \quad\quad\quad\quad\quad\quad\quad\quad\;\;\;\; + \|  \mathrm{x}_{i,t}(N) \|_{P_i}, \label{cost function}  \allowdisplaybreaks \\
 &\text{s.t.} \;  \mathrm{x}_{i,t}  (0)   =  x_{i,t}, \label{cons1}   \allowdisplaybreaks \\
 & \quad\;\, \mathrm{x}_{i,t}(\tau+1)  = A_i \mathrm{x}_{i,t}  (\tau) + B_i \mathrm{u}_{i,t}  (\tau),   \label{local dynamic constraint} \allowdisplaybreaks \\
 & \quad\;\,  (  \mathrm{x}_{i,t} (\tau)  ,   \mathrm{u}_{i,t}  (\tau)  ) \in  \mathcal{C}_{x,i} \times \mathcal{C}_{u,i},   \label{stage input constraint} \allowdisplaybreaks \\
 & \quad\;\,  \mathrm{x}_{i,t} (N)   \in  \mathcal{C}_{f,i},  \label{terminal constraint} \\
 & \quad\;\, \tau =  0, \cdots, N-1, \; \forall  i  \in  \mathcal{M}. \nonumber
\end{align}
\end{subequations}  
{The local decision variables include the sequences $\textbf{x}_{i,t} = [\mathrm{x}_{i,t}^\top(0), \allowbreak \cdots, \allowbreak \mathrm{x}^\top_{i,t}(N)]^\top$
and $\mathbf{u}_{i,t} \allowbreak  = \allowbreak  [\mathrm{u}^\top_{i,t}(0), \allowbreak \cdots, \allowbreak  \mathrm{u}^\top_{i,t}(N-1)]^\top$.
Let $\mathrm{x}_{\mathcal{N}_{i},t}(\tau) \allowbreak = \allowbreak [\mathrm{x}^\top_{i,t}(\tau),\allowbreak \cdots,\allowbreak \mathrm{x}^\top_{j,t}(\tau)]^\top$ and $\mathrm{u}_{\mathcal{N}_{i},t}(\tau) \allowbreak = [\mathrm{u}^\top_{i,t}(\tau), \allowbreak \cdots, \allowbreak \mathrm{u}^\top_{j,t}(\tau)]^\top$ be the concatenated decision variables of the agents in $\mathcal{N}_i$ for $\tau = 0,\cdots,N-1$.
The global decision variables include the sequences $\textbf{X}_t =[\mathrm{X}^{\top}_{t}(0),\cdots, \mathrm{X}^{\top}_{t}(N)]^\top$ and $\mathbf{U}_t \allowbreak  = \allowbreak  [\mathrm{U}^{\top}_{t}(0), \allowbreak  \cdots, \allowbreak  \mathrm{U}^{\top}_{t}(N-1)]^\top$,
where $\mathrm{X}_{t}(\tau) = [\mathrm{x}_{1,t}^\top(\tau),\cdots,\mathrm{x}_{M,t}^\top(\tau)]^\top$ for $\tau = 0,\cdots,N$ and $\mathrm{U}_{t}(\tau) \allowbreak = \allowbreak[\mathrm{u}_{1,t}^\top(\tau), \allowbreak \cdots, \allowbreak \mathrm{u}_{M,t}^\top(\tau)]^\top$ for $\tau = 0,\cdots,N-1$.
Let $\textbf{X}^*_t$ and $\mathbf{U}^{*}_t$ be the optimal solution.}
Local terminal cost $\| \cdot \|_{P_i}$ and stage cost with coupling $ \| \cdot \|_{H_{i}}$ are considered, 
with $H_i,P_i \in \mathbb{S}_{++}$.
The convex sets $\mathcal{C}_{x,i}$, $\mathcal{C}_{u,i}$, and $\mathcal{C}_{f,i}$ represent the local state, input, and terminal constraint sets, respectively.
{We define matrices $Q, R, P \in \mathbb{S}_{++}$, which satisfy $\operatorname{blkdiag}(Q,R) = \sum_{i \in \mathcal{M}} {E_{i}^\top H_i E_{i}}$ and $P = \operatorname{blkdiag}(\allowbreak P_1, \allowbreak \cdots, \allowbreak P_M)$.}

\begin{assumption}
{
Consider the MAS \eqref{global dynamics}. Let $l(X,U) = \| X \|_Q + \| U \|_R$, $l_{f}(X) \allowbreak = \allowbreak \|  X \|_{P}$, and $\mathcal{C}_{f} \allowbreak = \allowbreak \mathcal{C}_{f,i} \allowbreak \times \allowbreak \cdots \allowbreak \times \allowbreak \mathcal{C}_{f,M}$.
There exist local stabilizing control laws $K_{f,i} x_i$ such that the MAS \eqref{global dynamics} under $\pi_{f}(X) := [(K_{f,i} x_i)^\top,\cdots,(K_{f,M} x_M)^\top]^\top$ satisfies, $\forall X \in \mathcal{C}_{f}$:}

{
\noindent $\bullet \;\; A X + B\pi_{f}(X)\in \mathcal{C}_{f}\textup{ and } \pi_{f}(X)\in \mathcal{C}_{u,i} \times \cdots \times \mathcal{C}_{u,M}$,}

{
\noindent $\bullet \;\; l_{f}(A X + B\pi_{f}(X))-l_{f}(X) \leq - l(X, \pi_{f}(X))$.
}
\end{assumption}
 
{The DMPC problem \eqref{DMPC problem} can be formulated as \eqref{problem:PDproblem}, 
with local and global decision variables $z_{i,t} = [\textbf{x}^{\top}_{i,t},\mathbf{u}^\top_{i,t} ]^\top$ and $z_{t}\allowbreak = \allowbreak [\textbf{X}^{\top}_{t}, \allowbreak \mathbf{U}^\top_{t} ]^\top$, respectively.
The local and global cost functions are $f_i(z_{\mathcal{N}_i,t}) \allowbreak =  \allowbreak \|  \mathrm{x}_{i,t}(N)  \|_{P_i} \allowbreak +  \allowbreak \sum_{\tau=0}^{N-1}   \| [ \mathrm{x}^\top_{\mathcal{N}_i,t}(\tau) , \mathrm{u}^\top_{\mathcal{N}_i,t}(\tau) ]^\top \|_{H_{i}}$
and $\Scale[1]{f(z_t) = \|z_t \|^2_H = 
\sum_{i \in \mathcal{M}} f_i( z_{\mathcal{N}_i, t}) }$, respectively, 
where $H = \operatorname{blkdiag}(I_{N -1} \otimes {\operatorname{blkdiag}(Q,R)}, P)$, with $\otimes$ the Kronecker product.
The local constraint set $\mathcal{C}_i \allowbreak = \allowbreak \mathcal{C}_{x,i} \allowbreak \times \allowbreak \mathcal{C}_{u,i} \allowbreak \times \allowbreak \cdots \times \mathcal{C}_{f,i}$ is convex and satisfies Assumption 1.
Since $H_i,P_i \in \mathbb{S}_{++}$, Assumption 2 is satisfied, with $L_i = 2\max(\overline{\lambda}(H_i),\overline{\lambda}(P_i))$.
Thus, \eqref{DMPC problem} formulated as \eqref{problem:PDproblem} can be solved using Alg. 1 with the convergence guarantee in \eqref{subbound}.
Let $z_t^*=[\textbf{X}^{*\top}_{t}, \allowbreak \mathbf{U}^{*\top}_{t} ]^\top$ and $z_t^K=[\textbf{X}^{K\top}_{t}, \allowbreak \mathbf{U}^{K\top}_{t} ]^\top$ be the optimal and $K_{\text{th}}$-iteration sub-optimal solutions, respectively. The corresponding optimal and sub-optimal control laws are
\begin{align}
  \quad  \pi^*(X_t)   & := \mathrm{U}^*_t(0) = [{\mathrm{u}^{* \top}_{i,t}}(0),  \cdots,    {\mathrm{u}^{* \top}_{M,t}}(0)]^\top  = \Xi  z^*_t, \label{opt control} \\
   \pi^K(X_t)   & := \mathrm{U}^K_t(0) =    [{\mathrm{u}^{K \top}_{i,t}}(0), \cdots,   {\mathrm{u}^{K \top}_{M,t}}(0)]^\top = \Xi  z^K_t,  \label{sub control}
\end{align}
respectively, where $\Xi \in \mathbb{R}^{{ m_u} \times {m_z}}$ is a selection matrix. 
Let the optimal cost of \eqref{DMPC problem} be $\Scale[0.99]{V(X_t) := f(z^*_t)  = \|z^*_t \|^2_H}$.}

\begin{assumption}
Consider the DMPC problem \eqref{DMPC problem}, there exists a Lipschitz constant $L$ such that for two feasible initial states $X_{t_1}$ and $X_{t_2}$, it holds that
\begin{equation} \label{existence of L}
    \|  z^*_{t_1} - z^*_{t_2} \| \leq L \|  X_{t_1} - X_{t_2} \|,
\end{equation}
where $z^*_{t_1}$ and $z^*_{t_2}$ are the optimal solutions of \eqref{DMPC problem} with initial states $X_{t_1}$ and $X_{t_2}$, respectively. 
\end{assumption}

\begin{remark}
{Since \eqref{DMPC problem} is a convex optimization problem parametrized by $X_t$, $L$ always exists \cite{hager}.}
When all the constraints in \eqref{DMPC problem} are polytopic, the Lipschitz constant $L$ exists and can be determined as the maximal gain of the explicit MPC solution of \eqref{DMPC problem} in {\cite{LipEstimate}}.
When the DMPC problem \eqref{DMPC problem} only contains input constraints, the Lipschitz constant $L$ can be determined analytically {\cite{dominic1}}.
{For more general problem setups, sampling-based methods can be used to estimate $L$ with specified probability guarantee \cite{lip_estimate}.}
\end{remark}


\subsection{DMPC Framework with Quantization Refinement}

{Let $T$ be the given communication data rate, defined as the number of bits available per time step for one agent to transmit to another agent.
The communication constraint $n K \leq T$ exists and limits the number of iterations $K$ that Alg. 1 can be implemented and introduces quantization noise into Alg. 1.}
{We assume sufficient computation resources are provided for carrying out all computations.}
To address the communication constraint, we propose a DMPC framework with a quantization refinement scheme in Alg. 2, which contains an \textbf{off-line stage} for quantization parameter design and an \textbf{on-line stage} for quantization refinement and control input computation.

In the \textbf{off-line stage}, we assume all agents have access to the global information.
{Given data rate $T$, the quantization bit number $n$ and iteration number $K$ satisfying the communication constraint $n K \leq T$ are determined (Step 1).
Then, the the global state $X_0$ is measured for computing the optimal solution $z^*_0$ (Step 2). 
For initialization (Step 3), the warm-start solution $z^K_0$ is set to $z^*_0$ and the base quantization intervals $C_{i,t}$ are set to $0$.}

\begin{algorithm}[t]
\caption{DMPC Framework with Quantization Refinement}
\begin{algorithmic}[0]
    \State \textbf{\textup{Require:}} $T$, $L$, $(1-\gamma)<\rho<1$, $ \eta<\frac{1}{L_f}$, $\gamma = \frac{\alpha_f}{L_f}$, {$t = 0$};
    \State \textbf{Off-line Stage:} 
\State 1. Determine $n$ and $K$ satisfying $n K \leq T$;
\State 2. Measure $X_0$ and compute the optimal solution $z^*_0$ of \eqref{DMPC problem};
\State 3. {Initialize $z^{K}_{0} = z^{*}_0$ and $C_{i,0} = 0$, $i \in \mathcal{M}$.}
    \State \textbf{On-line Stage:} 
    \If{$t = 0$}
        \State 4. {Set $\hat{X}_{i,0} = X_0, i \in \mathcal{M}$ and apply optimal control $\pi^*(X_t)$};
    \Else
        \State 5. {Obtain $\hat{X}_{i,t}$ and compute $\Delta \hat{X}_{i,t-1} := \hat{X}_t - \hat{X}_{i,t-1}$, \\ \hspace*{2.55em}$i \in \mathcal{M}$;}
        \State 6. {Update base quantization interval $C_{i,t}$ using \eqref{update rule a}, $i \in \mathcal{M}$;}
        \State 7. Update initial solution $z^0_{\mathcal{N}_i,t} \leftarrow z^{K}_{\mathcal{N}_i,t-1}$, {$i \in \mathcal{M}$};
        \State 8. Formulate DMPC problem \eqref{DMPC problem} as a distributed optimiza \\ \hspace*{2.55em}-tion problem \eqref{problem:PDproblem} with $x_{i,t}$ as the time-varying parameter \\\hspace*{2.55em}and solve it using Alg. 1 over the MAS \eqref{global dynamics} for $K$ \\ \hspace*{2.55em}iterations with inputs $n$, $\rho$, $\eta$, $z^0_{i,t}$, {$C_{i,t}$, $i \in \mathcal{M}$}; 
        \State 9. Apply sub-optimal {control $\pi^K(X_t)$};
    \EndIf
    \State 10. {Increase time step }$t = t + 1$;
\end{algorithmic}
\end{algorithm}

{In the \textbf{on-line stage}, at every time step $t$, we require each agent $i$ to obtain a local estimate $\hat{X}_{i,t}$ of the global state $X_t$.
The estimate $\hat{X}_{i,t}$ consists of accurate measurements $x_{i,t},i\in\mathcal{N}_i$ (also used as initial states for the DMPC problem \eqref{DMPC problem}) and estimates $\hat{x}_{j,t},j\in \mathcal{M} \backslash \mathcal{N}_i$.
Let $e_{i,t} := \hat{X}_{i,t} - X_t$ be the combined estimation error.
When $t=0$, the local estimates $\hat{X}_{i,0}$ are set as $X_0$ from Step 2 and the optimal control $\pi^K(X_t)$ is applied to each agent.
When $t > 0$, the agents compute an estimated global state change $\Delta \hat{X}_{i,t-1} := \hat{X}_{i,t} - \hat{X}_{i,t-1}$ (Step 5) and update the local base quantization intervals (Step 6) with
\begin{align} 
     C_{i,t} =    b_1 C_{i,t-1} + b_2 \|{\Delta} \hat{X}_{i,t-1}\|   + (b_3 + b_4 b_5) e, \label{update rule a}  
\end{align} 
where $e \geq 0$ is an upper-bound of $e_{i,t}$ for all $t > 0$, and
\begin{equation} \label{b_1_4_definition}
\begin{aligned} 
 &b_1 := \frac{ \rho^K  ( 1 +  M a_1 a_3  - M  a_2)}{1 -   M a_2 }, b_2  :=  \frac{L a_1 }{1 -    M a_2},  b_5  :=  \frac{  4 b_2 }{ 1-b_1},\\
&b_3 := \frac{ 4 M a_2 b_2   +   2 L  a_1}{1-M a_2}, b_4  :=  \frac{ M a_2 (  b_1  -  \rho^K ) +\rho^K M   a_1  a_3 }{1-  M a_2 }. 
\end{aligned}
\end{equation}
Then, the previous solutions are used as warm-start solutions (Step 7) for solving the current DMPC problem \eqref{DMPC problem} by implementing Alg. 1 over the MAS \eqref{global dynamics} for $K$ iterations (Step 8).}

\begin{assumption}  \label{assump estimation error}
{The errors satisfy $\max_{i \in \mathcal{M}, t \geq 0} (e_{i,t}) \leq e$.}
\end{assumption}

\begin{remark}
    {The estimates $\hat{X}_{i,t}$ can be obtained using, e.g., quantized distributed state estimation algorithms \cite{distributed1}, \cite{distributed2}.
    To satisfy Assumption 5, we can reduce the estimation error by reducing the quantization level and increase iteration number of these algorithms.} 
\end{remark}

{The update rule \eqref{update rule a} contains a refinement component and an adaption component.
The former corresponds to $b_1 C_{i,t-1}$, which allows $C_{i,t-1}$ to decrease (if $b_1 < 1$).
The later corresponds to $b_2 \|{\Delta} \hat{X}_{i,t-1}\| + (b_3 + b_4 b_5) e$, which allows $C_{i,t-1}$ to increase and adapt to the global state change $\Delta X_{t-1} := X_t - X_{t-1}$. 
The term $(b_3 + b_4 b_5) e$ preemptively increases $C_{i,t-1}$ to compensate for the estimation error $e_{i,t}$.
With appropriate $n$ and $K$, the update rule \eqref{update rule a} and the warm-start step form a time-step-wise quantization refinement scheme, which allows the quantization intervals to be refined over time, as illustrated in Fig. 1. 
This on-line scheme is key to our stability analysis (Proposition 3 and Theorem 1).}


\section{Stability Analysis of the Proposed Method} \label{sec stability}

{In this section, we derive sufficient conditions on $n$ and $K$ for guaranteeing recursive feasibility and closed-loop stability of the MAS \eqref{global dynamics} controlled by the proposed DMPC framework in Alg. 2.}

\subsection{Recursive Feasibility Guarantee} \label{recur}
{We prove recursively feasible of MAS \eqref{global dynamics} controlled by Alg. 2, i.e., $X_t$ is a feasible for \eqref{DMPC problem} $\forall t > 0$.
To solve \eqref{DMPC problem} at Step 8 of Alg. 2 with convergence guarantee, we also need to ensure Lemma 2 holds.
This requires the intervals $C_{i,t-1}$ determined by \eqref{update rule a} satisfy \eqref{C condition} for $\forall t > 0$.
In Proposition 1, we prove these two conditions hold simultaneously.}

\begin{proposition}
Consider the MAS \eqref{global dynamics} controlled by the DMPC framework in Alg. 2 with a feasible state $X_0$.
Suppose Assumptions 1-5 hold.
Let Alg. 2 be initialized with $T$, $L$, $\rho \in (1-\gamma, 1)$ and $\eta<\frac{1}{L_f}$.
{If the quantization bit number $n$ satisfies \eqref{n definition}, and
\begin{align} \label{n definition 2}
    1 - M a_2 > 0,
\end{align}
where $a_2$ in \eqref{a_1_3 definitions} depends on $n$, then $X_t$ is recursively feasible.}

\begin{proof}
{We first provide some auxiliary results.
When $z^*_{t-1}$ and $z^*_{t}$ exist and the warm-start in Step 7 of Alg. 2 is applied, we have
\begin{align} 
 & \|  {\Delta z^{0}_{t}} \| := \| z^0_{t} - z^*_{t} \|  = \| z^{K}_{t-1} - z^*_{t-1}  + z^*_{t-1} - z^*_{t}  \|   \allowdisplaybreaks  \nonumber \\
      &  \leq \| z^{K}_{t-1} - z^*_{t-1}  \| + \| z^*_{t-1} - z^*_{t} \|   \overset{\eqref{existence of L}}{\leq} \|  {\Delta z^{K}_{t-1}} \| + L \| \Delta X_{t-1} \|. \label{rho to epsilon bound} 
\end{align}
From Assumption \ref{assump estimation error}, we have
\begin{align} 
   &  \| \Delta \hat{X}_{i,t-1} -  \Delta X_{t-1} \|  =   \| \hat{X}_{i,t} - \hat{X}_{i,t-1} -  X_{t} + X_{t-1} \| \\
\leq & \| \hat{X}_{i,t} - X_{t}  \| + \|  X_{t-1} - \hat{X}_{i,t-1} \| = e_{i,t} + e_{i,t-1} \leq  2e, \label{dx_dxa_bound}
\end{align}
which implies
\begin{align} \label{dx_dxa_bound}
  \Scale[1]{\| \Delta X_{t-1} \| \leq \| \Delta \hat{X}_{i,t-1} \| + 2e, \| \Delta \hat{X}_{i,t-1} \| \leq    \| \Delta X_{t-1} \|  + 2e.}
\end{align}
By definition of the update rule \eqref{update rule a}, we have 
\begin{align} 
   &  C_{\Sigma,t+1}  =   b_1 C_{\Sigma,t} + b_2 \sum_{i\in\mathcal{M}} \| \Delta \hat{X}_{i,t} \| +  M b_3 e   +  M  b_4 b_5 e \label{update rule sum}\\
   & \overset{\eqref{dx_dxa_bound}}{\leq} b_1 C_{\Sigma,t} + M b_2 \| \Delta X_{t} \| + 2 M b_2 e + M b_3 e   + M b_4 b_5 e \label{update sum equality} \\
  & \overset{\eqref{dx_dxa_bound}}{\leq}  b_1 C_{\Sigma,t} + b_2 M   \| \Delta \hat{X}_{i,t} \|  + 4 M b_2 e  + M b_3 e   + M b_4 b_5 e. \label{update sum bound}
\end{align}
Since \eqref{n definition 2} holds, $b_1$-$b_4$ in \eqref{update rule a} have positive values.}

{We prove Proposition 1 by induction.
Let $t = 1$ be the base case and $t = g > 1$ be the induction case.
The induction assumptions are \textbf{(i)} $X_{t-1}$ being feasible for \eqref{DMPC problem}, \textbf{(ii)} $a_1 \|  {\Delta z^{0}_{t-1}} \| + a_2 C_{\Sigma,t-1}  \leq  C_{i,t-1}$, and \textbf{(iii)} $\| C_{i,t-1} -   C_{\Sigma,t-1} / M \|  \leq b_5 e $.
Showing \textbf{(i)} hold $\forall t > 0$ gives us recursive feasibility, while \textbf{(ii)} and \textbf{(iii)} are necessary for guaranteeing Alg. 1 has convergence guarantee \eqref{subbound} $\forall t > 0$.}

{Base case: When $t = 1$, we construct a shifted input sequence 
$\mathbf{U}^s_{1} = [{\mathrm{U}^{* \top}_{0}}(1),\cdots, {\mathrm{U}^{* \top}_{0}}(N), \pi_f( \mathrm{X}^*_{0}(N))^\top]^\top$
and state sequence
$\mathbf{X}^s_{1} = [\mathrm{X}^{* \top}_{0}(1),\cdots, \mathrm{X}^{* \top}_{0}(N), (A \mathrm{X}^*_{0}(N) + B \pi_f (\mathrm{X}^*_{0}(N)))^\top]^\top$ from the optimal input sequence 
$\mathbf{U}^*_{0} \allowbreak =  \allowbreak [{\mathrm{U}^{* \top}_{0}}(0),  \allowbreak \cdots,  \allowbreak {\mathrm{U}^{* \top}_{0}}(N-1)]^\top$
and state sequence 
$\mathbf{X}^*_{0}  \allowbreak =  \allowbreak [{\mathrm{X}^{* \top}_{0}}(0),  \allowbreak \cdots,  \allowbreak {\mathrm{X}^{* \top}_{0}}(N)]^\top$ obtained at $t = 0$.
Since $X_1 = A X_0 + B \mathrm{U}^*_0(0) = \mathrm{X}^*_{0}(1)$ and $\pi_f$ defined in Assumption 3 guarantees positive invariance of $\mathrm{X}^{*}_{0}(N)$ in $\mathcal{C}_f$, $\mathbf{U}^s_{1}$ and $\mathbf{X}^s_{1}$ form a feasible solution of \eqref{DMPC problem}. This implies $X_1$ is feasible for \eqref{DMPC problem}.
Since $X_0$ is feasible for \eqref{DMPC problem}, \eqref{rho to epsilon bound} holds.
Applying \eqref{rho to epsilon bound} and \eqref{update sum bound}, and using the facts $\|  {\Delta z^{K}_{0}}  \| = 0$ and $C_{\Sigma,0} = 0$, we obtain
\begin{align}
 &  a_1 \|  {\Delta z^{0}_1}  \| + a_2 C_{\Sigma,1} \nonumber \allowdisplaybreaks\\
 \leq & a_1 ( L \| \Delta \hat{X}_{i,0} \| + 2 L e)  \nonumber \allowdisplaybreaks\\
  &   + a_2 ( b_2 M   \| \Delta \hat{X}_{i,0} \| + ( M b_2  4 + M b_3 )e   + M b_4 b_5 e )  \allowdisplaybreaks\\
 = & ( a_1 L   + a_2 b_2 M ) \| \Delta \hat{X}_{i,0} \|  + ( a_2  ( 4 M b_2 + M b_3 )+  2 L a_1  ) e  \nonumber \allowdisplaybreaks\\
 & + M a_2 b_4 b_5 e   \allowdisplaybreaks \\
 \leq & b_2 \| \Delta \hat{X}_{i,0} \| + b_3 e + b_4 b_5 e = C_{i,1},  \label{base case bound}
\end{align}
where \eqref{base case bound} holds by the definition of the update rule \eqref{update rule a}.
To show $\| C_{i,1} -  C_{\Sigma,1} / M \|  \leq b_5 e$, we use \eqref{update rule a} and \eqref{update sum equality} to obtain
\begin{align}
    & \| C_{i,1} - C_{\Sigma,1}/M \| \leq  \|   b_1 C_{i,0} + b_2 \| \Delta \hat{X}_{i,0}\| + b_3 e + b_4 b_5 e  \allowdisplaybreaks \nonumber  \allowdisplaybreaks \\
&\;\;\;\;\;\;\;\;\; -( b_1 C_{\Sigma,0}/ M + b_2 \| \Delta X_{0} \|  + (2 b_2 + b_3) e + b_4 b_5 e ) \| \allowdisplaybreaks \\
&\hspace{-3pt}\leq b_2 \|   \| \Delta \hat{X}_{i,0}\| -     \| \Delta X_{0} \| \|    + 2 b_2 e  \leq b_2 \|   \Delta \hat{X}_{i,0} -  \Delta X_{0} \|   + 2 b_2 e  \label{induction bound inter 0} \allowdisplaybreaks \\
&{\leq} 4 b_2  e \leq b_5 e,  \label{base case bound 2}
\end{align}
where we used the reverse triangle inequality in \eqref{induction bound inter 0}.
The inequalities in \eqref{base case bound 2} hold due to {\eqref{dx_dxa_bound}} and the definition of $b_5$, respectively.}

{Induction case: When $t = g > 1$, from induction assumptions \textbf{(i)} and \textbf{(ii)}, we know $X_{g-1}$ is feasible for \eqref{DMPC problem} and $a_1 \|  {\Delta z^{0}_{g-1}}  \| \allowbreak + \allowbreak a_2 C_{\Sigma,g-1} \allowbreak \leq \allowbreak  C_{i,g-1}$.
Thus, \eqref{DMPC problem} can be solved using Alg. 1 with convergence guarantee \eqref{subbound} at $t = g-1$ to obtain feasible sequences $\mathbf{X}^{K}_{g-1}$ and $\mathbf{U}^{K}_{g-1}$. 
Using $\mathbf{X}^{K}_{g-1}$, $\mathbf{U}^{K}_{g-1}$, and $\pi_f$, we construct shifted sequences $\mathbf{U}^s_{g}$ and $\mathbf{X}^s_{g}$, like $\mathbf{U}^s_{1}$ and $\mathbf{X}^s_{1}$ in the base case, which form a feasible solution of \eqref{DMPC problem} and guarantee $X_g$ is feasible for \eqref{DMPC problem}. Thus, \eqref{rho to epsilon bound} holds.
Applying \eqref{rho to epsilon bound} and \eqref{update sum bound} gives
\begin{align}
 &  a_1 \|  {\Delta z^{0}_t} \| + a_2 C_{\Sigma,g} \nonumber \\
 \leq & a_1 (  \|  {\Delta z^{K}_{g-1}} \| + L \| \Delta \hat{X}_{i,g-1} \| + 2 L e)  + a_2 b_1 C_{\Sigma,g-1}  \nonumber \\
  &   + a_2 (   b_2 M   \| \Delta \hat{X}_{i,g-1} \| + ( M b_2  4 + M b_3 )e   + M b_4 b_5 e )  \\
 = & a_1 \|  {\Delta z^{K}_{g-1}} \| + a_2 b_1 C_{\Sigma,g-1} + ( a_1 L   + a_2 b_2 M ) \| \Delta \hat{X}_{i,0} \|    \nonumber \\
 & + ( a_2  ( 4 M b_2 + M b_3 )+  2 L a_1  ) e + M a_2 b_4 b_5 e.   \label{induction assumption bound 0} 
\end{align}
From induction assumption \textbf{(ii)}, $\|  {\Delta z^{0}_{g-1}} \|$ can be bounded as:
\begin{align} \label{induction assumption}
 \|  {\Delta z^{0}_{g-1}} \| \leq  C_{i,g-1} / a_1 - a_2/a_1 C_{\Sigma,g-1}.
\end{align}
Since $n$ satisfies \eqref{n definition}, \eqref{subbound} holds and we can obtain
\begin{align} 
   \|  {\Delta z^{K}_{g-1}} \| &  \leq  \rho^K (\| {\Delta z^{0}_{g-1}} \| + a_3 C_{\Sigma,g-1}) \\
   \overset{\eqref{induction assumption}}{\leq} & \rho^K ( (C_{i,g-1} / a_1 - a_2/a_1 C_{\Sigma,g-1}) + a_3 C_{\Sigma,g-1}). \label{induction assumption bound}
\end{align}
Then, bounding $\|  {\Delta z ^{K}_{g-1}}\|$ in \eqref{induction assumption bound 0} with \eqref{induction assumption bound} gives
\begin{align}
 &   a_1 \|  {\Delta z^{0}_g} \| + a_2  C_{\Sigma,g}  \leq   \rho^K a_1   (C_{i,g-1} / a_1 - a_2/a_1 C_{\Sigma,g-1}) \nonumber\\
 & \quad\; + \rho^K a_1 a_3 C_{\Sigma,g-1} + a_2 b_1 C_{\Sigma,g-1} + ( a_1 L   + a_2 b_2 M ) \| \Delta \hat{X}_{i,0} \|    \nonumber \\
 &\;\;\,\;\; + ( a_2  (  4 M b_2 + M b_3 )+  2 L a_1  ) e + M a_2 b_4 b_5 e.   \label{induction bound 4}
\end{align}
From induction assumption \textbf{(iii)}, i.e., $\| C_{i,g-1} - C_{\Sigma,g-1}/M \| \leq b_5 e$, we derive $ \| C_{\Sigma,g-1}/M \|  \leq  \| C_{i,g-1} \| + b_5 e$, which we use to obtain
\begin{align} \label{induction bound 5}
    C_{\Sigma,g-1} = M (C_{\Sigma,g-1}/M) {\leq} M C_{i,g-1} + M b_5 e.
\end{align}
Then, applying \eqref{induction bound 5} to \eqref{induction bound 4} yields the following {after carrying out simplification}:
\begin{align}
 &  a_1 \|  {\Delta z^{0}_g}  \| + a_2 C_{\Sigma,g} \nonumber \\
 &\leq    ( \rho^K  (1 - M a_2  + M a_1 a_3) + M a_2  b_1 )  C_{i,g-1}   \nonumber \\
 & \; + (a_1  L +  a_2  b_2 M)   \| \Delta \hat{X}_{i,g-1} \|  +   ( M a_2  b_2  4 + M a_2  b_3 + 2 L a_1 ) e \nonumber \\
 & \; + M b_5 ( a_2 b_4 + a_2  b_1 - \rho^K a_2  +  a_1  \rho^K a_3)   e   \\
&   \leq  b_1 C_{i,g-1} + b_2 \| \Delta \hat{X}_{i,g-1} \|   +  b_3 e + b_4 b_5 e = C_{i,g} \label{induction bound 7}
\end{align}
where \eqref{induction bound 7} holds by definition of the update rule \eqref{update rule a}.
To show $\| C_{i,g} - C_{\Sigma,g}/M \|  \leq b_5 e$, we use \eqref{update rule a} and \eqref{update sum equality} to obtain
\begin{align}
    &\| C_{i,g-1} - C_{\Sigma,g-1}/M \| \nonumber \\
\leq  & \| (  b_1 C_{i,g-1}  -  b_1 C_{\Sigma,g-1}/M ) \nonumber \\
& \qquad\qquad\;\;\;\;  + ( b_2 \| \Delta \hat{X}_{i,g-1}\| - b_2    \| \Delta X_{g-1} \|) + 2 b_2 e \|   \label{induction bound inter 3} \\
\leq & b_1 \| C_{i,g-1}  - C_{\Sigma,g-1}/M \| + b_2 \|   \Delta \hat{X}_{i,0} - \Delta X_{0} \|  + 2 b_2 e \label{induction bound inter 4}\\
\leq & b_1 b_5 e + 4 b_2 e \leq b_5 e, \label{induction bound inter 5}
\end{align}
where we used the reverse triangle inequality in \eqref{induction bound inter 4}.
The two inequalities in \eqref{induction bound inter 5} hold due to the induction assumption $\| C_{i,g-1}  - C_{\Sigma,g-1}/M \| \leq b_5 e$ and the definition of $b_5$, respectively.}

{Finally, combining the base and induction cases shows \textbf{(i)}-\textbf{(iii)} hold $\forall t > 0$, where \textbf{(i)} gives us recursive feasibility.}
\end{proof}
\end{proposition}


\subsection{Interconnected System with Three Subsystems}

\begin{figure}[t!]
 \centering
\includegraphics[width=0.78\hsize]{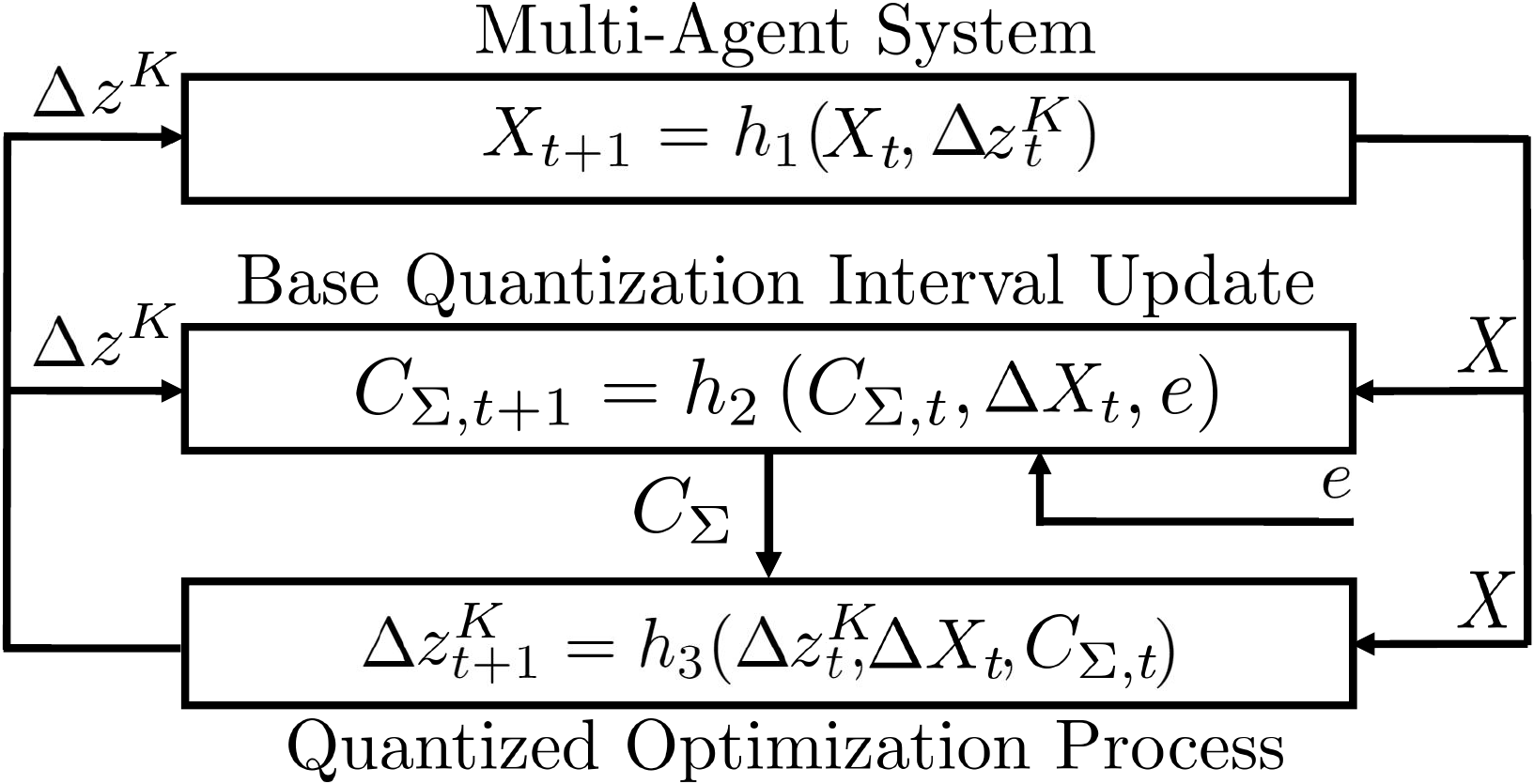}
\caption{The interconnection of the three subsystems.}
\label{interconnection}
\end{figure}

{In order to carry out stability analysis using the small-gain theorem, we propose to consider the MAS \eqref{global dynamics} controlled by Alg. 2 as three interconnected subsystems, including the sub-optimally controlled MAS, the base quantization interval update process, and the quantized optimization process, as shown in Fig. 2.
Their states are the state of the MAS $X_t$, the sum of base quantization interval $C_{\Sigma,t}$, and the sub-optimality of the optimization algorithm $\Delta z^K_t$, respectively.}
{The parameter $e$ from \eqref{update rule a} enters Subsystem 2 as an external input.}
\vspace{5pt}

\noindent \textbf{Subsystem 1.} Given the DMPC problem \eqref{DMPC problem} and Alg. 2, the dynamics of the sub-optimally controlled MAS is formulated as
\begin{align}
    X_{t+1} & =  h_1(X_t, \Delta z^K_t) \nonumber \\
    & := A X_t + B  \Xi z^K_t = A X_t + B  \Xi z^*_t + B \Xi \Delta z^K_t, \label{sub1}
\end{align}
{where $B \Xi \Delta z^K_t$ represents a virtual perturbation resulted from the sub-optimality associated with $z^K_t$.}
\vspace{5pt}

\noindent {\textbf{Subsystem 2.} The dynamics of the base quantization interval update process in Step 5 and 6 of Alg. 2 is formulated as
\begin{align}
  C_{\Sigma, t+1} & = h_2(C_{\Sigma, t},\Delta X_{t}, e) \nonumber \\
 & :=   b_1 C_{\Sigma,t} + b_2 \sum_{i\in\mathcal{M}} \| \Delta \hat{X}_{i,t} \| + M ( b_3  +  b_4 b_5)e,  \label{sub2}
\end{align}
which corresponds to the sum of updates in \eqref{update rule sum}.}
\vspace{5pt}

 \noindent \textbf{Subsystem 3.} The dynamics of the quantized distributed optimization process formed by Step 5-8 of Alg. 2 is {formulated as
\begin{align} \label{sub3}
   \Delta z^K_{t+1}   = h_3(  \Delta z^K_t, \Delta X_{t},  C_{\Sigma,t}),
\end{align}
which corresponds the sub-optimality bound in \eqref{subbound} for time step $t + 1$.}
\vspace{5pt}

{Next, we identify ISS-Lyapunov functions for the subsystems, and derive conditions on $n$ and $K$ for satisfying the small-gain conditions \eqref{small gain conditions original} and guaranteeing stability of the interconnected system.}

\subsection{ISS-Lyapunov Functions for Subsystem 1, 2, and 3}

{Inspired by \cite{dominic1}, we prove $\psi(X_t) := \sqrt{V(X_t)}$ is an ISS-Lyapunov function for Subsystem 1. 
We also show $C_{\Sigma,t}$ and $\|  {\Delta z^{K}_t} \|$ are ISS-Lyapunov functions for Subsystems 2 and 3, respectively, if $n$ and $K$ satisfy certain conditions.
We present proofs of Propositions 2-4, and Theorem 1 in the Appendix.}

\begin{proposition} [\textbf{ISS-Lyapunov Function for Subsystem 1}]
Consider Subsystem 1 in \eqref{sub1} with an feasible initial state $X_0$.
{Suppose Assumptions 1-5 and Proposition 1 hold.}
Then, $\psi(X_{t})$ is an ISS-Lyapunov function of \eqref{sub1} satisfying
\begin{equation} \label{ISS sys1}
     \psi(X_{t+1})  \leq  (1-\alpha_1 ) \psi(X_{t})  +  \gamma_{1,3} \|  {\Delta z^{K}_t} \|,
\end{equation}
with $\alpha_1 := 1- \sqrt{ 1 - { \underline{\lambda} (Q) }/({ L^2 \overline{\lambda} (H)} })$ and $\gamma_{1,3} := \sqrt{\overline{\lambda} (H)} \| B \| L$.
\end{proposition}

\noindent The difference between Proposition 2 and Theorem 1 in \cite{dominic1} is we define the Lipschitz constant $L$ differently in \eqref{existence of L}. The bound in Lemma 3 on $\Delta X_t$ is used in the proofs of Propositions 3 and 4.

\begin{lemma}[Lemma 6, \cite{dominic2}] 
Consider the MAS \eqref{global dynamics} controlled by $\pi^K(X_t)$ defined in \eqref{sub control}.
The change in the state $\Delta X_{t} := X_{t+1} - X_t$ of the MAS satisfies
\begin{align} \label{delta x simplify}
 \| \Delta X_{t}\|   \leq  c_1 \psi(X_t)   + \| B  \| \| {\Delta z^{K}_t} \|,
\end{align}
where $c_1 := ({\| A - I\|   + \| B\|}){\sqrt{\underline{\lambda}(H)}}$.
\end{lemma}

\begin{proposition} [\textbf{ISS-Lyapunov Function for Subsystem 2}]
Consider Subsystem 2 in \eqref{sub2}.
{Suppose Assumptions 1-5 and Proposition 1 hold.
If $n$ and $K$ satisfy 
\begin{align} \label{sys K1 condition}
   K \geq  K_1  := \left\lceil - \log_{\rho}\left(\frac{1 -   M a_2 }{   M a_1 a_3  + 1 - M  a_2 }\right) \right\rceil,
\end{align}
where $a_1$-$a_3$ in \eqref{a_1_3 definitions} depend on $n$, then $C_{\Sigma,t}$ is an ISS-Lyapunov of \eqref{sub2} satisfying
\begin{align} \label{final iss 3}
      {  C_{\Sigma,t+1}   \leq (1-\alpha_2  ) C_{\Sigma,t}   + \max ( \gamma_{2,1}  \psi(X_t) , \gamma_{2,3}  \|  {\Delta z ^{K}_t}\| , \gamma^e_{2}  e ) ,}
\end{align}
where $\alpha_2  := 1 - b_1$, $\gamma_{2,1}  := 3 b_2 M  c_1$, $\gamma_{2,3}  :=  3  M  \| B  \| b_2$, and $\gamma^e_{2}  :=  3 M ( 4 b_2  + b_3 + b_4 b_5)$.}
\end{proposition}

\begin{proposition} [\textbf{ISS-Lyapunov Function for Subsystem 3}]
Consider Subsystem 3 in \eqref{sub3}.
{Suppose Assumptions 1-5 and Proposition 1 hold.
If $n$ and $K$ satisfy 
\begin{align}  \label{sys2 K condition}
K \geq  K_2  := \left\lceil - \log_{\rho}\left(c_2\right) \right\rceil,
\end{align}
where $c_2 := {1}/({ 1  +  L \| B  \|  + a_3  M  \| B  \| b_2 })$, with $a_3$ in \eqref{a_1_3 definitions} and $b_2$ in \eqref{b_1_4_definition} depending on $n$,
then $\|  {\Delta z^{K}_t} \|$ is an ISS-Lyapunov function of \eqref{sub3} satisfying
\begin{align}  \label{ISS function 2}
  \|  {\Delta z^{K}_{t+1}} \|  \leq  (1 - \alpha_3  ) \|  {\Delta z^{K}_t} \|  +  \max ( \gamma_{3,1}  \psi(X_t)  , \gamma_{3,2}  C_{\Sigma,t}, \gamma^e_{3} e ),
\end{align}
where $\alpha_3  := 1 - \rho^{2K} / b_1$, $\gamma_{3,1} := 3 \rho^K (  L + M b_2 ) c_1$, $\gamma_{3,2}  := 3 \rho^K a_3 b_1$, and $\gamma^e_{3} :=  \gamma^e_{2}$.}
\end{proposition}

\subsection{Conditions on the Quantization Parameters for Stability} \label{main theorem subsection}

{We now derive conditions on $n$ and $K$, given $T$, such that the three interconnected cycles in Fig. 2 satisfy the small-gain conditions \eqref{small gain conditions original} which further implies stability of the interconnected system.
Let $\mathcal{X}_{i,j} := \tilde{\alpha}_{i}^{-1} \circ \left(\mathrm{Id}-\tilde{\mu}_{i}\right)^{-1} \circ \tilde{\gamma}_{i,j}, \; i \in \{ 1,2,3\}$,
with 
\begin{align*}
    & \tilde{\alpha}_1 (s)  := \alpha_1 s, \;\;\;\;\; \tilde{\alpha}_2 (s)  := \alpha_2 s,  \;\;\;\;\; \tilde{\alpha}_3 (s) := \alpha_3 s, \\
    & \tilde{\gamma}_{1,3} (s)   :={\gamma}_{1,3} s,   \; \tilde{\gamma}_{2,1} (s)  :={\gamma}_{2,1} s, \; \tilde{\gamma}_{2,3} (s) := {\gamma}_{2,3} s, \\
    &   \tilde{\gamma}_{3,1} (s) :={\gamma}_{3,1}  s,  \; \tilde{\gamma}_{3,2} (s) :={\gamma}_{3,2}  s,
\end{align*}
the small-gain conditions corresponding to the three cycles are: there exist $\tilde{\mu}_{i}, i \in \{ 1,2,3\}$ such that}
\begin{subequations} \label{cycle conditions 123}
  \begin{align}
    \mathcal{X}_{1,3} \circ \mathcal{X}_{3,1} &< \text{Id}, \allowdisplaybreaks \\ \mathcal{X}_{2,3} \circ \mathcal{X}_{3,2} &< \text{Id}, \allowdisplaybreaks  \\
    \mathcal{X}_{1,3} \circ \mathcal{X}_{3,2} \circ \mathcal{X}_{2,1} & < \text{Id}.  
\end{align}  
\end{subequations}

\begin{theorem}
Consider the MAS \eqref{global dynamics} controlled by the DMPC framework in Alg. 2.
Suppose Assumptions 1-5 hold.
{Given communication data rate $T$ defined as the number of bits that can be transmitted per time step.}
{If the quantization bit number $n$ and optimization iteration number $K$ satisfy $n K \leq T$, \eqref{n definition}, \eqref{n definition 2}, and
\begin{align} 
    K  \geq \max(K_1,K_2,K_3,K_4,K_5), \label{final condition on K}
\end{align}
where $K_1$ defined \eqref{sys K1 condition}, $K_2$ defined in \eqref{sys2 K condition}, and 
\begin{subequations} 
\begin{flalign}
    &    K_3 :=  {\left \lceil \operatorname{log}_{\rho} \left( \frac{   \alpha_1 - 3 M {\gamma}_{1,3}  b_2 c_1}{ 3 L   {\gamma}_{1,3} c_1 +   \alpha_1 c_2) }  \right)  \right \rceil},   \allowdisplaybreaks \\
    &   K_4 := \left \lceil \operatorname{log}_{\rho} \left( \frac{-\Bar{b}_1+\sqrt{ \bar{b}^2_1 - 4 \Bar{a}_1 \Bar{c}_1 }}{ 2 \Bar{a}_1}   \right) \right \rceil,  \allowdisplaybreaks \\
  &    K_5 := \left \lceil \operatorname{log}_{\rho} \left(  \frac{-\Bar{b}_2+\sqrt{ \bar{b}^2_2 - 4 \Bar{a}_2 \Bar{c}_2 }}{ 2 \Bar{a}_2} \right)  \right \rceil,  \allowdisplaybreaks 
\end{flalign}
\end{subequations}
depend on $n$, with
\begin{align*}
    &\Bar{a}_1 := \frac{   {\rho^K}  c_2 - {b_1}}{  {3 {\rho^K} \gamma}_{2,3} a_3 },  \Bar{b}_1 := \frac{   -  {\rho^K}  - {b_1} c_2    }{  {3  {\rho^K} \gamma}_{2,3} a_3  }, \Bar{c}_1 := \frac{  {b_1}  }{  {3 {\rho^K} \gamma}_{2,3} a_3  },  \\
    & \Bar{a}_2 := \frac{   {\rho^K} \alpha_1  c_2 }{3 {\gamma}_{1,2} {\gamma}_{2,3} {b_1}  c_1 },  \Bar{b}_2 :=  \frac{  -  {\rho^K}  \alpha_1 -  \alpha_1 {b_1}  c_2 - L{b_1}}{3 {\gamma}_{1,2} {\gamma}_{2,3} {b_1}  c_1 } , \\
    & \Bar{c}_2 :=\frac{  \alpha_1  }{3 {\gamma}_{1,2} {\gamma}_{2,3}  c_1 } - M b_2,
\end{align*}
then the MAS \eqref{global dynamics} controlled by Alg. 2 is recursively feasible} {and the closed-loop system is ISS w.r.t the parameter $e$ in \eqref{update rule a} considered as a constant external input.}
\end{theorem}

\begin{remark}
    {If $e = 0$ as a result of, i.e. all agents having accurate estimates of the global state (the network has a central node), the interconnected system becomes asymptotically stable.}
\end{remark}

\begin{remark}
    {The small-gain conditions \eqref{cycle conditions 123} naturally embed a degree of conservatism.
\cite{jiang_necessary} proved small-gain-type conditions as necessary (tight) and sufficient for a family of interconnected systems. 
Extending this result to our problem setting can be a future direction.
Bad estimation of Lipschitz constants may also lead to conservatism in the bound \eqref{final condition on K} since the value of $\max(K_1, \allowbreak K_2, \allowbreak K_3, \allowbreak K_4, \allowbreak K_5)$ depends on $L_i$, $L_f$, and $L$.
To reduce $L$, pre-conditioning the optimization problem underlying \eqref{DMPC problem} is viable \cite{dominic1}.
At the cost of potentially reduced closed-loop performance, one can also tune the DMPC problem parameters (e.g., $H_i$ and $P_i$ in \eqref{cost function}) and adjust the network topology (e.g. reducing degree $d$ of $\mathcal{G}$) to reduce $L_i$, $L_f$, and $L$.}
\end{remark}

\begin{remark}
{Given a limited data rate $T$, it is possible a combination of $n$ and $K$ satisfying $n K \leq T$, \eqref{n definition}, \eqref{n definition 2}, and \eqref{final condition on K} does not exist.
In this case, one can solve the quantization parameter design problem:
	\begin{align} \label{problem: min datarate}
    	T_{\min} = \underset{ n, K \in \mathbb{Z}_+}{{\min}} n K, \text{ s.t. }  \eqref{n definition}, \eqref{n definition 2}, \eqref{final condition on K}.
 	\end{align}
to determine the minimum data rate $T_{\min}$ required to implement Alg. 2 with closed-loop stability guarantee.
Note that \eqref{problem: min datarate} is solvable off-line through enumeration.}
\end{remark}


\section{Multi-AUV Formation Control Problem} \label{sec design}

\begin{figure}[t!]
 \centering
\includegraphics[width=1\hsize]{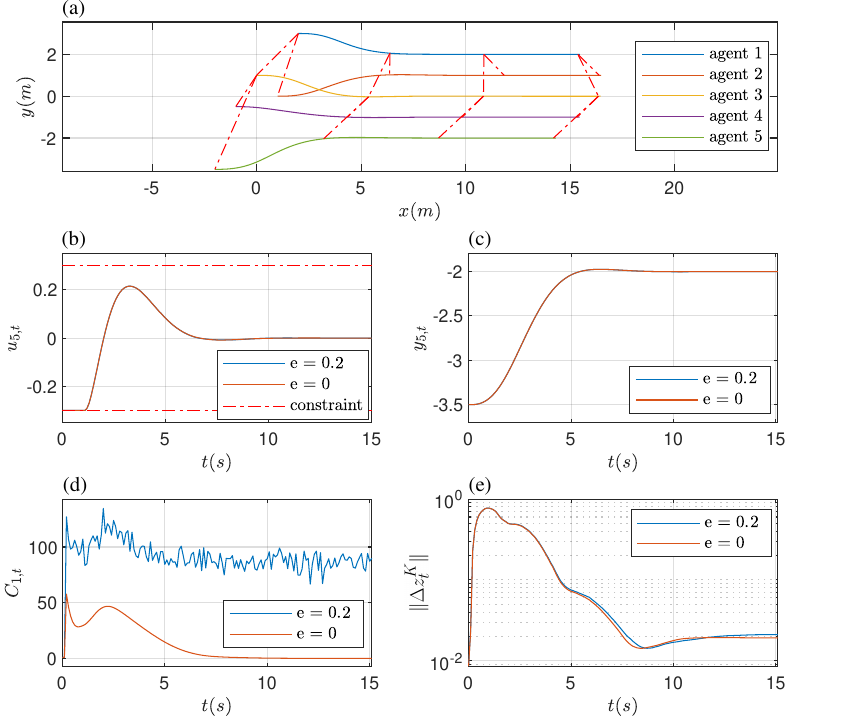}
\caption{(a) the trajectories of the AUVs; (b) the input $u_{5,t}$ of AUV 5; (c)-(e) the change in $y_{5,t}$, $C_{1,t}$, and $\|  {\Delta z^{K}_t} \|$ w.r.t. time step $t$.}
\label{interconnection}
\end{figure}

We demonstrate the proposed algorithm and the theoretical findings using a control example with multiple autonomous underwater vehicles (AUVs), which suffers from limited underwater communication data rate.
There are five AUVs moving in the $xy$-plane (horizontal) and communicating according the neighbourhoods $\mathcal{N}_1 = \{1,2,3\}$, $\mathcal{N}_2 = \{1,2\}$, $\mathcal{N}_3 = \{1,3,4\}$, $\mathcal{N}_4 = \{3,4\}$, and $\mathcal{N}_5 = \{3,5\}$.
The MAS is controlled by a $y$-axis controller based on Alg. 2, which we focus on, and a low-level $x$-axis controller. 
The AUV dynamics in the $y$-axis is $\zeta_{i,t+1} = A_i \zeta_{i,t} + B_i u_{i,t}$, where $\zeta_{i} := [y_{i} ,\delta_{i}, \omega_{i}]^\top$ and
\begin{equation}
    A_i =
        \begin{bmatrix}
        1 & T_d v_c & 0 \\
        0 & 1 & T_d\\
        0 & 0 & 1 - {N_{\omega}}/{ \mathcal{I}}
    \end{bmatrix}, \;
    B_i =
    \begin{bmatrix}
        0 \\
        0\\
       {1}/{\mathcal{I}} 
    \end{bmatrix},
    \nonumber
\end{equation}
with position $-5 \leq  y_i \leq 5$ m, yaw angle $-1 \leq \delta_i \leq 1$ rad, angular velocity $ -2 \leq \omega_i \leq 2$ rad/s, input torque $ -0.3 \leq u_i \leq 0.3$ N$\cdot$m/s, velocity $v_c = 1$ m/s, rotational inertia $\mathcal{I} = 2$ kg$\cdot$m\textsuperscript{2}, damping coefficient $N_{\omega} \allowbreak = \allowbreak 1$ N$\cdot$m/(rad/s), and sampling time $T_d = 0.1$ s.
{This constraint setup satisfies Assumption 1.}
The initial states of the AUVs are
$[y_{1,0},\allowbreak y_{2,0},\allowbreak y_{3,0},\allowbreak y_{4,0},\allowbreak y_{5,0}]^\top = [5,\allowbreak 0,\allowbreak 1,\allowbreak -0.5,\allowbreak -3.5\allowbreak ]^\top$ m,
$\delta_{i,0} = 0$ rad, and $\omega_{i,0} = 0$ rad/s, $\forall i \in \{ 1,\cdots,5\}$.
Their reference states are
$[y_{r,1},\allowbreak  y_{r,2}, y_{r,3},\allowbreak y_{r,4},\allowbreak y_{r,5}]^\top  =\allowbreak [ 2,\allowbreak 1,\allowbreak 0,\allowbreak -1,\allowbreak -2]^\top $ m,
$\delta_{r,i} = 0$ rad, and $\omega_{r,i} = 0$ rad/s, $\forall i \in \{ 1,\cdots,5\}$.
The DMPC formulation in Section \textup{III-A} of \cite{ACC2021} is implemented, which imposes penalty on the relative $y$-distance errors between AUVs $i$ and $j$, $\forall j \in \mathcal{N}_i$. 
{Choosing $N = 18$, $Q_i, \allowbreak R_i, \allowbreak S_{ij} \in I$, $P_i$, $K_{f,i}$ satisfying Assumption 3 is determined by solving the algebraic Ricatti inequality and enforcing appropriate constraints on $P$ to ensure a separable structure.
Local terminal constraint $\mathcal{C}_{f,i} := \{ \zeta_i \mid \| \zeta_i\|^2_{P_i} \leq 0.057 \}$ are determined through ellipsoidal approximation. 
Since the local cost functions depend $Q_i$, $S_i$, $R_i$, $P_i \in \mathbb{S}_{++}$, Assumption 2 is satisfied.}

To obtain a Lipschitz constant {satisfying Assumption 4}, we uniformly sample global states $\zeta_m$ and inputs $u_m$ from the constraints and compute $\tilde{\zeta}_m = A \zeta_m + B U_m$.
We collect 2000 pairs of $(\zeta_m, \tilde{\zeta}_m)$ that are both feasible for the DMPC problem.
The Lipschitz constant $L = 9.72$ is computed from $L = \max_{1 \leq m \leq 2000} \frac{\| \zeta_m - \tilde{\zeta}_m \|}{\| z^*(\zeta_m) - z^*(\tilde{\zeta}_m) \|}$. 
{Given $T = 32$ kB/s, the parameters for Alg. 2 are chosen as $\rho = 0.986$, $\eta = 0.001$, $n  = 26$, and $K = 1000$, where $n$ and $K$ satisfy conditions \eqref{n definition}, \eqref{n definition 2}, and \eqref{final condition on K}.}

{We simulate and compare two cases: with and without estimation noise. 
For the first case, we sample $e_{i,t}$ uniformly from $[-0.05,0.05]$ and set $e = 0.2$ to satisfy Assumption 5.
For the second case, we set $e = 0$.}
{Fig. 3(a) shows trajectories of the AUVs controlled by Alg. 2 with $e = 0.2$, where the red dashed lines represent the current formation. It can be seen the AUVs converge to the references while forming the desired formation.
Fig. 3(b) shows the input $u_{5,t}$ of AUV 5, where the input constraint $-0.3$ N$\cdot$m/s is activated before $t = 1.3$ s.}
Fig. 3(c) shows the convergence of the position $y_{5,t}$ of AUV 5 (part of the state of Subsystem 1) to the reference state $y_{r,5} = -2$ m.
{Fig. 3(d) shows the evolution of $C_{1,t}$ (part of the state of Subsystem 2), which converges to a neighbourhood of $0$ when $e = 0.2$ and to $0$ when $e = 0$.
Fig. 3(e) shows $\| {\Delta z^{K}_t} \|$ (normed state of Subsystem 3) converges to a plateau near $10^{-2}$ in both cases, which results from limited solver accuracy in the projection step of Alg. 1.
After $t = 5$ s, the value of $\|  {\Delta z^{K}_t} \|$ with $e = 0.2$ is slightly higher than that with $e = 0$.}
{As can be seen, the effect of the parameter $e$ is most obvious on the base quantization interval (Subsystem 2) and least visible in the state (Subsystem 1).
This is expected since $e$ enters Subsystem 2 as a constant external input and its effect indirectly propagates through Subsystem 3 to Subsystem 1, as illustrated in Fig. 2.}
{To conclude, the simulation results demonstrated that for a given data rate $T$, if the quantization bit number $n$ and iteration number $K$ satisfy the conditions in Theorem 1, then, the interconnected system formed by \eqref{sub1}-\eqref{sub3} is ISS w.r.t. the parameter $e$ considered as an external input.}

\section{Conclusions}
We proposed a novel real-time DMPC framework with a quantization refinement scheme for MASs with limited communication data rates and derived sufficient conditions on the quantization parameters for guaranteeing the closed-loop stability.
{Future works can focus on quantitative methods for jointly designing the DMPC formulations (e.g., cost functions) and the algorithm parameters (e.g., $n$, $K$) to reduce the required communication data rate for guaranteeing closed-loop stability while maintaining certain level of performance.}
{Another direction is to address the conservatism resulted from bounding $e_{i,t}$ with $e$,
by developing an ISS-Lyapunov function for the estimation error $e_{i,t}$ w.r.t. the change in state $\Delta X_t$.
This would enable us to consider the state estimation dynamics as another subsystem interconnected with the subsystems in \eqref{sub1}-\eqref{sub3}, and apply the small-gain theorem to prove asymptotically stable of the overall system.}


\section*{Appendix}


\subsection{Proof of Proposition 2} \label{p2 proof}
{From Proposition 1, $X_t$ is feasible and $\pi^K(X_t)$ and $\pi^*(X_t)$ exist $\forall t \geq 0$.}
Let $X_{t+1} = A X_t + B \pi^K(X_t)$, $\breve{X}_{t+1} = A X_t + B \pi^*(X_t)$, $\psi(X_{t+1}) := \|z^*_{t+1}\|_H$, $\psi(\breve{X}_{t+1}) := \|\breve{z}^*_{t+1}\|_H$, we know that
\begin{equation} \label{t2_1}
  \psi(X_{t+1})  \leq |  \|z^*_{t+1}\|_H - \|
  \breve{z}^*_{t+1}\|_H | +  \|\breve{z}^*_{t+1}\|_H .
\end{equation}
From Assumption 3, we have 
\begin{align}  \label{p2 inter1}
      V(\breve{X}_{t+1}) \leq V(X_{t}) - \| X_{t} \|^2_Q.
\end{align}
Lower bounding $\| X_{t} \|^2_Q$ with
\begin{equation} 
   \| X_{t} \|^2_Q \geq \underline{\lambda} (Q) \| X_{t}\|^2 \overset{\eqref{existence of L}}{\geq} \frac{\underline{\lambda} (Q)}{L^2}  \| z^*_{t} \|^2 \geq  \frac{ \underline{\lambda} (Q) }{ L^2 \overline{\lambda} (H)} \| z^*_{t} \|^2_H, \nonumber
\end{equation}
and then taking square-roots on both sides of \eqref{p2 inter1} yields
\begin{equation} \label{l1_4}
    \psi(\breve{X}_{t+1}) \leq \sqrt{ 1 - { \underline{\lambda} (Q) }/({ L^2 \overline{\lambda} (H)}) }\psi(X_{t}).
\end{equation}
Applying the reverse triangle inequality, we have 
\begin{align}
  &  | \| z^*_{t+1} \|_H - \| \breve{z}^*_{t+1} \|_H |^2  \leq \| z^*_{t+1} - \breve{z}^*_{t+1}   \|^2_H  \allowdisplaybreaks \nonumber \\
    & \;\;\;\;\; \leq  \overline{\lambda}(H) \| z^*_{t+1} - \breve{z}^*_{t+1}   \|^2   \overset{\eqref{existence of L} }{\leq}  \overline{\lambda}(H) L^2 \| X^*_{t+1} - \breve{X}^*_{t+1}   \|^2.  \allowdisplaybreaks
\end{align}
Taking square-roots on both sides, we obtain
\begin{align}
 &   | \| z^*_{t+1} \|_H - \| \breve{z}^*_{t+1} \|_H |  \leq  \sqrt{\overline{\lambda}(H)} L \| X^*_{t+1} - \breve{X}^*_{t+1} \| \nonumber \\
    & \leq  \sqrt{\overline{\lambda}(H)} L \| B (\pi^*(X_t) - \pi^K(X_t) )  \|  \leq  \sqrt{\overline{\lambda}(H)} L \| B \| \|  {\Delta z^{K}_{t }} \|. \label{l2_4}
\end{align}
{Then, bounding the r.h.s of \eqref{t2_1} with \eqref{l1_4} and \eqref{l2_4} gives 
\begin{align}
  \psi(X_{t+1})  \leq  (1-\alpha_1) \psi(X_{t})  +  \gamma_{1,3} \|  {\Delta z^{K}_t} \|,  
\end{align}
where $\alpha_1 = 1- \sqrt{ 1 - { \underline{\lambda} (Q) }/({ L^2 \overline{\lambda} (H)})}$ and $\gamma_{1,3}  = \sqrt{\overline{\lambda} (H)} \| B \| L$.
Since $\| z^*\| \leq L\| X \|$ from \eqref{existence of L} and $ \| z^* \| = \|  [\textbf{X}^{* \top},\mathbf{U}^{* \top}]^\top \| \geq \| \mathrm{X}^*(0)\| = \| X \|$, we know $L \geq 1$.
Further, since $\overline{\lambda}(H) \geq \underline{\lambda}(Q)$, we know that $\alpha_1 \in (0,1)$. By definition, we also know $\gamma_{1,3} > 0$.}
Therefore, $\psi(X_{t})$ is an ISS-Lyapunov function for Subsystem 1. \hfill $\Scale[0.82]{\blacksquare }$

\subsection{Proof of Proposition 3} \label{p3 proof}
{Combining \eqref{dx_dxa_bound} with \eqref{delta x simplify} gives
\begin{align} \label{prop3 mid 0}
     \| \Delta \hat{X}_{i,t-1}\|   \leq  c_1 \psi(X_t)   + \| B  \| \|  {\Delta z^{K}_t}  \| + 2e.
\end{align}
Bounding $\| \Delta \hat{X}_{i,t-1}\|$ in \eqref{update sum bound} with \eqref{prop3 mid 0} and doing simplification give
\begin{align}
&  C_{\Sigma,t+1}  \leq b_1 C_{\Sigma,t} + M b_2  c_1 \psi(X_t)  \nonumber \\
  &   +  M  \| B  \| b_2 \|  {\Delta z^{K}_t} \| + M ( 4 b_2  + b_3 + b_4 b_5) e \label{prop 3 inter} \\
   & \leq (1-\alpha_2 )  C_{\Sigma, t}   + \max \left( \gamma_{2,1}  \psi(X_t) , \gamma_{2,3}  \|  {\Delta z ^{K}_{t}}\|, \gamma^e_{2} e \right). 
\end{align}
Since $\alpha_2 \in (0,1)$ from \eqref{sys2 K condition}
and $\gamma_{2,1}, \gamma_{2,3}, \gamma^e_2 > 0$ by definition,
$C_{\Sigma,t}$ is an ISS-Lyapunov function for Subsystem 2.}  \hfill $\Scale[0.82]{\blacksquare }$

\subsection{Proof of Proposition 4} \label{p4 proof}
{From Proposition 1, $n$ satisfies \eqref{n definition}, \eqref{subbound} holds, and we have
\begin{align}
     \|  {\Delta z^{K}_{t+1}} \| \overset{\eqref{rho to epsilon bound}}{\leq} & \rho^K ( (\|  {\Delta z ^{K}_{t}} \| + L \| \Delta X_{t} \|)  + a_3 C_{\Sigma,t+1}).
\end{align}
Then, bounding $\| \Delta X_{t} \|$ with \eqref{delta x simplify} and $C_{\Sigma,t+1}$ with \eqref{prop 3 inter} gives
\begin{align}
& \|  {\Delta z ^{K}_{t+1}}  \| \leq  \rho^K (1  +  L \| B  \|  + a_3  M  \| B  \| b_2 ) \|  {\Delta z ^{K}_t} \| + \rho^K a_3 b_1 C_{\Sigma,t} \nonumber \\
& \;\;\; +  \gamma_{3,1} (L + M b_2 ) c_1 \psi(X_t)   + M ( 4 b_2  + b_3 + b_4 b_5) e  \\
& \leq  (1 - \alpha_3  ) \|  {\Delta z ^{K}_t} \|   +  \max \left( \gamma_{3,1}  \psi(X_t), \gamma_{3,2}  C_{\Sigma, t}, \gamma^e_{3} e \right).
\end{align}
Since $\alpha_3 \in (0,1)$ from \eqref{sys2 K condition}
and $\gamma_{3,1}, \gamma_{3,2}, \gamma^e_3 > 0$ by definition,
$\|  {\Delta z^{K}_t}  \|$ is an ISS-Lyapunov function for Subsystem 3.} \hfill $\Scale[0.82]{\blacksquare }$

\subsection{Proof of Theorem 1} \label{final section}
Since the initial state $X_0$ is feasible and $n$ satisfies \eqref{n definition} and \eqref{n definition 2}, the MAS \eqref{global dynamics} under Alg. 2 is recursively feasible (Proposition 1). 

Since Assumptions 1-5 hold, the quantization bit number $n$ and iteration number $K$ satisfy \eqref{n definition}, \eqref{n definition 2}, and $K \geq \max(K_1,K_2)$, it is ensured Propositions 2-4 hold, i.e., Subsystems 1-3 admit \eqref{ISS sys1}, \eqref{final iss 3}, and \eqref{ISS function 2} as ISS-Lyapunov functions, respectively.

In addition, since the iteration number also satisfies $K \geq \max(K_3,K_4,K_5)$, the following conditions hold:
\begin{align*}
\tilde{\alpha}_{1}^{-1} \circ \tilde{\alpha}_{3}^{-1} \circ \tilde{\gamma}_{1,3} \circ \tilde{\gamma}_{3,1}  &<\textup{Id}, \\
\tilde{\alpha}_{2}^{-1} \circ \tilde{\alpha}_{3}^{-1} \circ \tilde{\gamma}_{2,3}\circ \tilde{\gamma}_{3,2}&<\textup{Id}, \\
\tilde{\alpha}_{1}^{-1} \circ \tilde{\alpha}_{2}^{-1} \circ \tilde{\alpha}_{3}^{-1} \circ \tilde{\gamma}_{1,3} \circ \tilde{\gamma}_{3,2} \circ \tilde{\gamma}_{2,1}  &<\textup{Id},
\end{align*}
which implies there always exist linear gain functions $\tilde{\mu}_{i}(s) := {\mu}_{i} s $, with $\mu_i > 0$, $i \in \{1,2,3 \}$, such that the small-gain conditions in \eqref{cycle conditions 123} hold, i.e., $\mathcal{X}_{1,3} \circ \mathcal{X}_{3,1} < \text{Id}$, $\mathcal{X}_{2,3} \circ \mathcal{X}_{3,2} < \text{Id}$, and $\mathcal{X}_{1,3} \circ \mathcal{X}_{3,2} \circ \mathcal{X}_{2,1} < \text{Id}$,
where $\mathcal{X}_{i,j} := \tilde{\alpha}_{i}^{-1} \circ \left(\mathrm{Id}-\tilde{\mu}_{i}\right)^{-1} \circ \tilde{\gamma}_{i,j}$.
{Therefore, the interconnected system formed by \eqref{sub1}-\eqref{sub3} is ISS w.r.t the parameter $e$ considered as a constant external input.} \hfill $\Scale[0.82]{\blacksquare }$

\bibliographystyle{IEEEtran}
\bibliography{IEEEabrv,MC}

\end{document}